# Rewards-based image analysis in microscopy

Kamyar Barakati[1, a], Yu Liu,[1] Utkarsh Pratiush[1], Boris N. Slautin[1] and Sergei V. Kalinin[1,3, b]

[1] Department of Materials Science and Engineering, University of Tennessee, Knoxville, TN 37996

[3] Pacific Northwest National Laboratory, Richland, WA 99354

## Abstract:

Imaging and hyperspectral data analysis is central to progress across biology, medicine, chemistry, and physics. The core challenge lies in converting high-resolution or high-dimensional datasets into interpretable representations that enable insight into the underlying physical or chemical properties of a system. Traditional analysis relies on expert-designed, multistep workflows, such as denoising, feature extraction, clustering, dimensionality reduction, and physics-based deconvolution, or on machine learning (ML) methods that accelerate individual steps. Both approaches, however, typically demand significant human intervention, including hyperparameter tuning and data labeling. Achieving the next level of autonomy in scientific imaging requires designing effective reward-based workflows that guide algorithms toward best data representation for human or automated decision-making. Here, we discuss recent advances in reward-based workflows for image analysis, which capture key elements of human reasoning and exhibit strong transferability across various tasks. We highlight how reward-driven approaches enable a shift from supervised black-box models toward explainable, unsupervised optimization on the examples of Scanning Probe and Electron Microscopies. Such reward-based frameworks are promising for a broad range of applications, including classification, regression, structure-property mapping, and general hyperspectral data processing.

[a] kbarakat@vols.utk.edu

[b] sergei2@utk.edu

## Introduction

Electron and Scanning Probe Microscopies are the foundational tools in materials science, condensed matter physics, chemistry, catalysis, electrochemistry, and other fields. Scanning Transmission Electron Microscopy (STEM) allows imaging and characterization of materials at nanoscale extending to atomic-resolution, providing insights into the atomic and molecular structures of materials. By now, STEM has become one of the primary tools in multiple academic and industrial research labs worldwide.[1-6] The versatility of STEM is further enhanced by its integration with techniques such as electron energy loss spectroscopy (EELS)[7-10] technique, which allows for precise analysis of the chemical composition,[11] electronic structure[12] of materials, and low energy quasiparticles.[13] The ability of STEM to image materials at nanometer and atomic levels makes it a crucial tool for advancing understanding of the structure-property relationship in a wide range of materials systems.[14-16] This combination is particularly beneficial in the development of new materials for technological applications, including semiconductors, solar cells[17], catalysts[18], and battery materials.[19]

Similarly, the extensive application of scanning probe microscopy (SPM) has opened the doors to exploring and modifying the nanoworld. Compared to other materials characterization tools, SPM offers a desktop footprint, low cost, and versatility in operating in multiple environments.[20, 21] It provides a wide range of functional imaging capabilities, extending from basic topographic imaging[22-26], to probing of electronic[23, 27], magnetic[22, 28, 29], mechanical[22, 23], biological[26, 30-34], and chemical[35, 36], properties. Furthermore, SPM supports multiple spectroscopy techniques in a variety of imaging modes, enabling comprehensive understanding and manipulation of materials at the nanoscale[37-39], and exploring phenomena such as single-molecule chemical transformation in biomolecules, polarization switching phenomena in ferroelectrics, and local electrochemical activity in a broad range of energy materials from batteries to fuel cells.[40, 41]

Microscopy tools now offer almost unbounded sources of information on the structure and property of matter in the form of images, spectra, and hyperspectral imaging data. Even for pure imaging, modern cameras and detectors offer capabilities up to 32k x 32k pixel images, well above the capability of human operator to examine. Similarly, high-dimensional data sets that are common across scanning probe and electron microscopies are generally outside the human ability to comprehend and interpret, necessitating lengthy iterative analyses. Currently, the efficiency of data utilization in microscopy is extremely low, often with 2-3 best images from the day of work being analyzed for publications of downstream applications. The data analysis itself often relies on the custom multistep workflows developed based on operator intuition and best practices and often take weeks to perfect. These factors lead to enormous hidden inefficiencies and strong operator biases in microscopy use. Compared to the success stories such molecular discovery by CryoEM[42], they also suggest the tremendous potential to increase efficiency and impact of imaging tool on broad range of scientific disciplines, if data analysis and potentially data acquisition methods are brought to the intrinsic data-generation limits of instruments.

The second set of opportunities relate to the real-time instrument operation. Any SPM and STEM operator is well familiar with the classical scan paradigm, and at some point, asked the question whether rectangular scanning and grid-based hyperspectral imaging orchestrated by human operator are indeed the only or the best way to explore new materials. The progress of big data methods in areas such as robotic vision[43] and autonomous driving[44] brings forth the question as to whether similar methods[45-48] can be useful for building automated microscopes. So far, these were predominantly realized in the form of workflows in which execution of the codes is driven by immediately available targets via fixed policies. For example, this can include the use of the deep convolutional networks or simpler image analysis tools for the identification of the *a priori* known object of interest such as atoms in scanning tunneling microscopy[49], identification of single DNA molecules[50], structural features in polymer crystallization[51, 52], spectroscopy of grain boundaries[53], and ferroelectric domain walls[54-58]. These developments are paralleled by the development of sampling methods such as compressed sensing.[59] More complex examples entail inverse workflows, in which the goal is to discover the structural features that maximize the desired aspect of the spectral response.[60] The number of examples of machine learning (ML) integration into active microscopy workflows has been growing rapidly over the last 5 years.[37, 61-70]

The rapid emergence of the application programming interfaces (APIs) now enables the direct implementation of ML workflows for automated microscopy. While 3 years ago these were generally lacking across microscopy manufacturers (with companies such as NION[71] in EM and Nanonis[72] in SPM being rare exceptions at that time), this trend is now changing drastically. For SPM, the API from SpecsGroup[73] is built for LabView[74] and with suitable translation layer allows integration with Python. The Python API from NanoSurf[75] offers direct communication to Python codes. On the user side, there have also been attempts to interface microscopy with ML and APIs. The AEcroscopy API[76, 77] developed at ORNL[78], and AESPM[79] developed by UT Knoxville[80] enable the full control of SPM manufactured by Asylum Research[81] but requires custom hardware. The DeepSPM API[82] based on the programming interface by SpecsGroup[73] provides APIs for Nanonis controllers. While the official APIs for SPMs are still scarce and come with limited control and data I/O for their microscopes, there is rapid trend to their emergence and operationalization, and congruent move from the customer community to build and share the in-house versions. The APIs have similarly become more advanced in the electron microscopy community over the last several years. Companies such as Thermo Fisher Scientific[83], Nion, and JEOL[84] have recently introduced APIs that allow for hardware control through code. The development of pure-software generic API that works universally on different microscopy can promote the development of automated microscopy,[85] but also requires development of the unsupervised image analysis pipelines.

Here we discuss recent advances in reward-based workflows for scientific imaging, which adopt key aspects of human expert decision-making and frame image analysis as a decision-making process in the space of possible operations. This formulation defines desired outcomes, establishes natural mappings onto classical decision-making approaches, and enables a high degree of transfer learning across tasks that may be very dissimilar. Reward-driven workflows shift image analysis away from supervised, hyperparameter-sensitive, and out-of-distribution-prone black-box models toward explainable, largely unsupervised, and robust optimization frameworks for constructing image-analysis pipelines. These workflows can be implemented as wrappers around both classical algorithms and Deep Convolutional Neural Network (DCNN)–based methods.

A central strength of this framework is that reward functions embed physical knowledge and expert heuristics directly into the optimization process by encoding experimentally meaningful criteria such as stability, contrast, structural consistency, and artifact suppression rather than generic performance metrics. This embeds physics into the search space itself, guiding optimization toward physically plausible and scientifically interpretable outcomes. The physics-based formulation is particularly well suited to real-time microscopy, where sparse data and low compute resources make large training datasets or high-fidelity simulators impractical. Integrating these physics-encoded rewards with probabilistic optimization enables uncertainty-aware decision-making, which is essential for safe and efficient active experimentation. Because the reward definitions rely on general physical principles rather than dataset-specific features, the resulting workflows remain robust across samples, materials systems, and operating conditions with minimal modification.

## I. Levels of automation in microscopy

To illustrate the opportunities for ML, we first discuss the levels of machine learning in microscopy based on the degree of integration of decision-making into the experiment planning (**Figure 1**). These levels describe a progression from human-driven analysis toward increasing algorithmic involvement in data interpretation and experimental control. Classically at the Level 1, image analysis during the experiments is based on human perception of images and spectra, with the operator choosing next action such as microscope tuning, selecting the region for imaging, or location for spectroscopic studies based on real time data. In this setting, ML plays no direct role in experiment execution, and the data stream is modified via a small number of pre- or dynamically configured hard-coded operations such as plane subtraction, contrast adjustment, etc. At the same time, more detailed analyses ranging from image segmentation to physics learning are performed after the experiment and correspondingly cannot affect experiment execution.

The second level of ML applications is the Level 2: real-time data analytics. Here, the data stream is processed in real time, and the generated insights are used to inform human decision making or can be used to realize the automated experiment workflows with the fixed policies, meaning that the sequence of experimental actions is predefined and not adaptively learned during execution. Generally, transition to real time analytics represents considerably more complex problems than post-acquisition analysis due to out-of-distribution drift. Even for simple analyses the human operator constantly tweaks image corrections and representations of the data. Segmentation workflows such as based on the deep convolutional networks suffer from out-of-distribution drift effects that materialize even for foundational models such as SAM (Segment Anything Model).[86] Real-time tuning of these algorithms is often challenging, since the training stage is often much more time-consuming than the inference stage. Hence, even for real-time analytics the key is the development of the fully unsupervised image analysis workflows that can represent the streaming high-D and often large, sampled data in the form that can inform human decision-making and instrument orchestration.

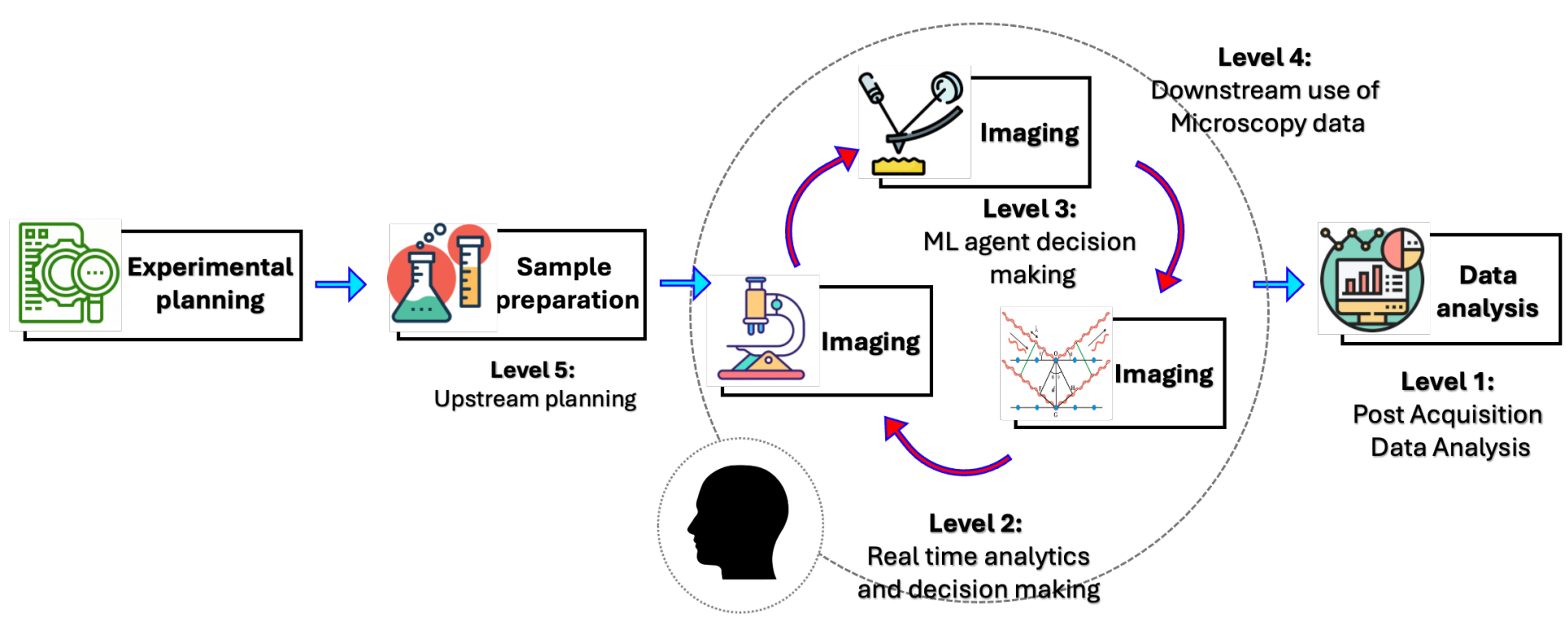

**Figure 1**: Levels of automation in microscopy.

This problem is even more acute for the Level 3 ML-enabled microscopy when the ML agent directly controls the instrument. At this level, algorithmic outputs are used to autonomously select experimental actions during acquisition. While many of these methods rely on active learning of the structure-property relationships, generally a more reliable segmentation method allows the use of simpler active learning policies. For example, exploration of functionalities at the topological or structural defects can be formulated as a fixed-policy experiment based on the segmentation workflow, whereas changes in image contrast can be used as feedback signals to execute image optimization stages.

Finally, the level 4 and 5 connect microscopy decision-making to general physics workflows, including downstream physics learning and upstream experiment design (e.g. sample making). These levels extend beyond individual experiments and integrate microscopy into broader, closed-loop scientific discovery pipelines. However, whether post-acquisition image analysis, real time analytics informing human operator, or autonomous decision-making, the rapid and reliable image analysis is a key requirement for microscopy experiments. Consequently, building analysis pipelines is the key element towards the progress.

## II. Current approaches for image analysis

### II.1. Classical image analysis methods

Image analysis has advanced significantly due to the increasing need for extracting meaningful information from complex visual data. Across multiple domains, from detecting macroscopic objects to identifying subtle anomalies indicative of diseases in biomedical imaging, or discovering previously unknown celestial bodies in astronomy, the demand for sophisticated analysis methods continues to grow.[87-89] As imaging technologies, such as high-resolution cameras and advanced microscopes, have evolved, so too has the necessity for robust computational approaches capable of processing and interpreting these vast datasets. The rapid progress in this field is driven by the development of more powerful algorithms, improved software frameworks, and machine learning techniques, enabling the extraction of insights from structures as minute as individual atoms in scanning transmission electron microscopy to large-scale environmental and astronomical observations captured by satellite imagery.[85]

Many image analysis techniques originated from well-established mathematical and computational methods designed to detect, quantify, and understand various features within an image. Preprocessing and enhancement are the first steps to prepare images for analysis. Filtering methods[90], such as Gaussian blur[91], median filtering[92], and sharpening[93], help remove noise and enhance key features. Histogram equalization[94] is used to improve image contrast, often aiding in the clearer visualization of medical images. Morphological operations[95] like erosion[96], dilation[97], opening, and closing[98] are applied to modify image structures, such as removing small objects or filling gaps, making images more suitable for further analysis. Edge detection[99] helps identify boundaries in an image by highlighting areas with significant intensity changes. Techniques like Sobel[100], Prewitt[101], and Canny[102] algorithms are commonly used for this purpose, making it easier to separate objects from the background or structural information for patterns and shapes analysis.

After preprocessing and edge detection, segmentation often becomes the next key step in analyzing image features. Thresholding[103] is a basic yet powerful technique that helps separate foreground objects from the background based on pixel intensity. Watershed segmentation[104] is another method that treats an image like a topographic map, making it particularly helpful for dividing overlapping structures, such as cells in biomedical images. Region growing[105] further enhances segmentation by grouping pixels with similar intensities, starting from seed points, often used to isolate organs or tissues. Segmentation techniques like these are essential to identify, quantify, and differentiate specific structures in an image, allowing for detailed and meaningful analysis. Once segmentation is complete, the next step is to extract important features from the segmented regions for further analysis.

Alternatively, keypoint detection[106] methods, such as Harris Corner Detector[107], SIFT (Scale-Invariant Feature Transform)[108], and SURF (Speeded-Up Robust Features)[109], help identify significant points that can be used for image matching. Feature descriptors[110], including Local Binary Patterns (LBP)[111] and Histogram of Oriented Gradients (HOG)[112], describe textures, shapes, and patterns, which are essential for tasks like classification and recognition.

Once key features are extracted, understanding the texture of these regions provides another layer of valuable information.[113] Gray-Level Co-Occurrence Matrix (GLCM)[114] calculates spatial relationships between pixel pairs, extracting features like contrast, correlation, and entropy, which are particularly useful in medical imaging and materials science. Fourier transform[115] analyzes frequency components to examine periodic textures, while wavelet transform[116] captures both spatial and frequency details, making it ideal for analyzing patterns that vary across an image. So far, this image analysis path has started with preprocessing to clean the image, followed by edge detection to find boundaries, and segmentation to separate distinct regions. Feature extraction and texture analysis provide details about the unique characteristics of these regions, leading to Object Detection and Recognition[117], which identifies meaningful components that align with human perception and hence can inform human decisions.

Template matching[118] uses a predefined template to identify similar areas within an image, commonly applied in industrial inspection to locate specific components or patterns. Hough transform[119] is used for detecting geometric shapes, like lines, circles, and ellipses, by mapping features to a parameter space, making it effective for structured shape recognition in various applications. The next step often involves understanding movement or aligning multiple images to make comparisons more meaningful, where Optical Flow techniques[120], such as Lucas-Kanade[121] and Horn-Schunck[122], estimate pixel motion across sequential images, which is useful for tracking object movement in video analysis and understanding scene dynamics. Image Registration[123] uses classical methods like mutual information[124] and cross-correlation[125] to align multiple images to a common coordinate system. This is particularly crucial in medical imaging, where registration is necessary to compare images taken at different times or using different imaging modalities.

With objects detected, features analyzed, and images registered, the next step is to categorize and identify patterns or anomalies within the data. K-Means Clustering[126] is a classical, unsupervised technique used to group similar pixels or features into clusters based on characteristics like color or intensity. Gaussian Mixture Models (GMM)[127] offer a probabilistic approach to clustering, allowing data points to belong to multiple clusters with different probabilities, which is useful when boundaries between clusters are not distinct. Other clustering methods, such as Hierarchical Clustering[128] and DBSCAN[129], provide additional options for grouping based on varying criteria, enhancing flexibility. Once the data is clustered, Anomaly Detection[130] helps identify unusual features that deviate from typical patterns. Classical methods like statistical outlier detection[131] or distance-based approaches[132] are used to detect these anomalies, which can be critical for quality control in manufacturing or for identifying abnormalities in medical images.

Classical image analysis involves structured steps, from preprocessing to segmentation, feature extraction, clustering, and anomaly detection. While effective, these methods rely on handcrafted features, predefined rules, and large real-time human involvement limiting their capacity in complex scenarios and especially for real-time image analysis.

### II.2. Deep learning

The advent of DCNNs[133] has transformed image analysis, allowing for automated feature learning directly from data. DCNNs enable more sophisticated tasks like classification, segmentation, and object detection by learning hierarchical features directly from data. DCNN architectures like AlexNet[134], VGG[135], ResNet[136], and EfficientNet[137] are commonly used for Image Classification, achieving state-of-the-art accuracy by learning a hierarchy of features—from simple edges to complex patterns—directly from large datasets, making them highly effective for diverse image recognition tasks.

The next step is to precisely outline and differentiate each component, which is where segmentation[138] plays a key role. Semantic Segmentation[139] assigns a class label to each pixel within an image, enabling the precise delineation of objects. Networks such as U-Net[140] and Fully Convolutional Networks (FCNs)[141] are commonly used for this task, making it particularly valuable in applications like medical imaging, where segmenting organs or tumors is crucial for diagnosis and treatment planning. Instance Segmentation[142] goes further by identifying each object instance individually. Models like Mask R-CNN[143] not only classify objects but also distinguish between different instances of the same class, allowing for more detailed analysis. This level of granularity is especially important in tasks where accurate count and separation of objects, such as cells or vehicles, are needed. Models like YOLO (You Only Look Once)[144], SSD (Single Shot MultiBox Detector)[145], and Faster R-CNN[146] are used for instance segmentation to provide both class labels and bounding box locations in object detection. Object detection is critical for applications such as autonomous driving, where identifying vehicles and pedestrians is necessary for safety, as well as in security systems and industrial automation for detecting components and monitoring environments.

Following segmentation and object detection, deep learning[147, 148] plays a crucial role in learning intricate features and refining images for further analysis. Keypoint detection models identify complex characteristic objects, such as facial landmarks or human joints, without relying on handcrafted features, enabling more detailed tasks like pose estimation and facial recognition. Autoencoders[149] and Variational Autoencoders (VAEs)[150], as unsupervised models, compress and reconstruct image data, making them effective for noise reduction and anomaly detection, thus enhancing image quality and interpretability. Image enhancement techniques like super-resolution[151] use models such as the Super-Resolution Convolutional Neural Network (SRCNN)[152] to map low-resolution images to high-resolution versions, offering clearer outputs. Additionally, Generative Adversarial Networks (GANs)[153] are employed to generate realistic images, perform style transformations, and enhance quality. CycleGAN[154], for instance, can transform images from one domain to another such as converting day scenes into night scenes making these approaches powerful for image synthesis and further enhancing image details.

Although DCNN models have revolutionized image analysis by learning complex features and enabling sophisticated tasks, they come with several significant limitations. They are largely supervised, requiring extensive labeled datasets, which is both time-consuming and costly to obtain. Additionally, these models are prone to overfitting, especially when trained on small datasets, and often struggle with generalization, meaning they may not perform well when faced with data that differs significantly from what they have seen during training—an issue known as out-of-distribution (OOD) shifts. Moreover, their interpretability remains a challenge, as the decision-making processes of deep models are often opaque. Lastly, their reliance on substantial computational resources can be a barrier to practical implementation in resource-constrained settings. These limitations emphasize the need for more efficient, adaptable, and interpretable image analysis techniques, keeping the quest for better image analysis methods wide open and an exciting area for further research.

**Table 1**: Partial list of image analysis tasks

| TASK NAME | DEFINITION | POPULAR ALGORITHMS |
|---|---|---|
| **IMAGE SEGMENTATION** | Dividing an image into multiple segments or regions to simplify analysis or focus on specific areas. | Classical: Watershed, Otsu's Thresholding, GrabCut<br>NN-based: U-Net, Mask R-CNN, DeepLab |
| **OBJECT DETECTION** | Identifying and locating objects within an image and labeling them with bounding boxes or masks. | Classical: Sliding Window, HOG + SVM<br>NN-based: YOLO, Faster R-CNN, SSD |
| **EDGE DETECTION** | Detecting boundaries or edges between different objects or regions within an image. | Classical: Sobel, Canny, Prewitt<br>NN-based: HolisticallyNested Edge Detection (HED) |
| **IMAGE CLASSIFICATION** | Assigning a label or class to the entire image based on its content. | Classical: k-NN, SVM<br>NN-based: ResNet, VGG, Inception, EfficientNet |
| **FEATURE EXTRACTION** | Identifying and extracting meaningful patterns or features from an image for further analysis. | Classical: SIFT, SURF, HOG<br>NN-based: CNN Feature Maps, Autoencoders |
| **IMAGE REGISTRATION** | Aligning two or more images, typically of the same scene taken at different times or angles. | Classical: RANSAC, Mutual Information, SIFT<br>NN-based: Spatial Transformer Networks (STN) |
| **IMAGE ENHANCEMENT** | Improving the visual quality of an image through processes like noise reduction or contrast adjustment. | Classical: Histogram Equalization, CLAHE<br>NN-based: SRCNN, GAN-based Enhancement |
| **IMAGE DENOISING** | Reducing noise from an image while preserving important details. | Classical: Gaussian Filter, Non-Local Means<br>NN-based: Denoising Autoencoders, DnCNN |
| **OPTICAL CHARACTER RECOGNITION (OCR)** | Detecting and converting text in an image into machine-readable format. | Classical: Tesseract, Template Matching<br>NN-based: CRNN, EAST Text Detector, Transformer-based OCR |
| **IMAGE RESTORATION** | Reconstructing or recovering an image that has been degraded, such as by blur or missing pixels. | Classical: Wiener Filter, Inverse Filtering<br>NN-based: GAN-based Inpainting, Deep Image Prior (DIP) |
| **SUPER-RESOLUTION IMAGING** | Enhancing the resolution of an image to reveal more details than the original image. | Classical: Bicubic Interpolation, Lanczos<br>NN-based: SRCNN, EDSR, ESRGAN |
| **CHANGE DETECTION** | Identifying differences between two or more images taken at different times or conditions. | Classical: Image Subtraction, Principal Component Analysis (PCA)-based[155] Change Detection<br>NN-based: Siamese Networks, FCNs |
| **PATTERN RECOGNITION** | Recognizing and categorizing patterns or shapes within an image, often using predefined templates. | Classical: Template Matching, k-NN, HOG<br>NN-based: CNN, Capsule Networks |
| **MOTION TRACKING** | Following and analyzing the movement of objects within a sequence of images or a video. | Classical: Kalman Filter, Optical Flow<br>NN-based: Recurrent Neural Networks (RNN), Track R-CNN |
| **COLOR ANALYSIS** | Analyzing the distribution and properties of colors in an image for segmentation or feature detection. | Classical: Histogram Analysis, K-means Clustering<br>NN-based: CNNs for Color Analysis |
| **IMAGE SMOOTHING** | Applying filters to an image to reduce sharp transitions and noise, making it appear softer. | Classical: Gaussian Blur, Median Filter<br>NN-based: CNN-based Autoencoders |
| **DEPTH ESTIMATION** | Estimating the distance of objects in an image from the camera, often used in stereo imaging. | Classical: Stereo Matching, Structure from Motion (SfM)<br>NN-based: MonoDepth, PSMNet |
| **MORPHOLOGICAL OPERATIONS** | Applying transformations such as dilation, erosion, opening, and closing to refine shapes in an image. | Classical: Dilation, Erosion, Opening, closing<br>NN-based: U-Net with Morphological Operations |
| **HISTOGRAM ANALYSIS** | Analyzing the distribution of pixel intensities within an image, often for thresholding or segmentation. | Classical: Histogram Equalization, Otsu's Method<br>NN-based: CNN-based Image Enhancement |
| **IMAGE FUSION** | Combining multiple images to produce a single image with enhanced information or clarity. | Classical: Wavelet Transform, PCA-based Fusion<br>NN-based: CNN-based Fusion, GAN-based Fusion |

Note that many DCNN architecture strictly speaking combine the supervised learning and the classical image analysis approaches. For example, instance segmentation (finding bounding boxes on object of interest) combines the NN classifier and non-maximal suppression principles, for example via Mask-RCNNs.

## III. From Tasks to workflows

The operations summarized in **Table 1** are examples of individual image analysis tasks. In practical applications, multiple tasks are performed sequentially, forming the image analysis workflow. An example can be denoising → keypoint detection → constructing descriptors → dimensionality reduction → visualization. Classically, the workflows are built by human including the selection of analysis operators and tuning hyperparameters of individual step. The value proposition of the deep learning-based analysis is that it allows to integrate multiple stages. For example, instead of manual feature engineering and building simple classifiers, the DCNN based classifiers can be trained on labeled data.

To illustrate these concepts further, an example of image analysis methods from Borodinov et al.[156] can be considered. Their approach involves a structured algorithm consisting of four key steps: (1) the selection of a structural descriptor to characterize key features of the image, (2) dimensionality reduction to process high-dimensional datasets and extract abundance maps, (3) the construction of a feature space for effective representation of the data, and (4) the assignment of image pixels to relevant structural types based on the generated feature space.

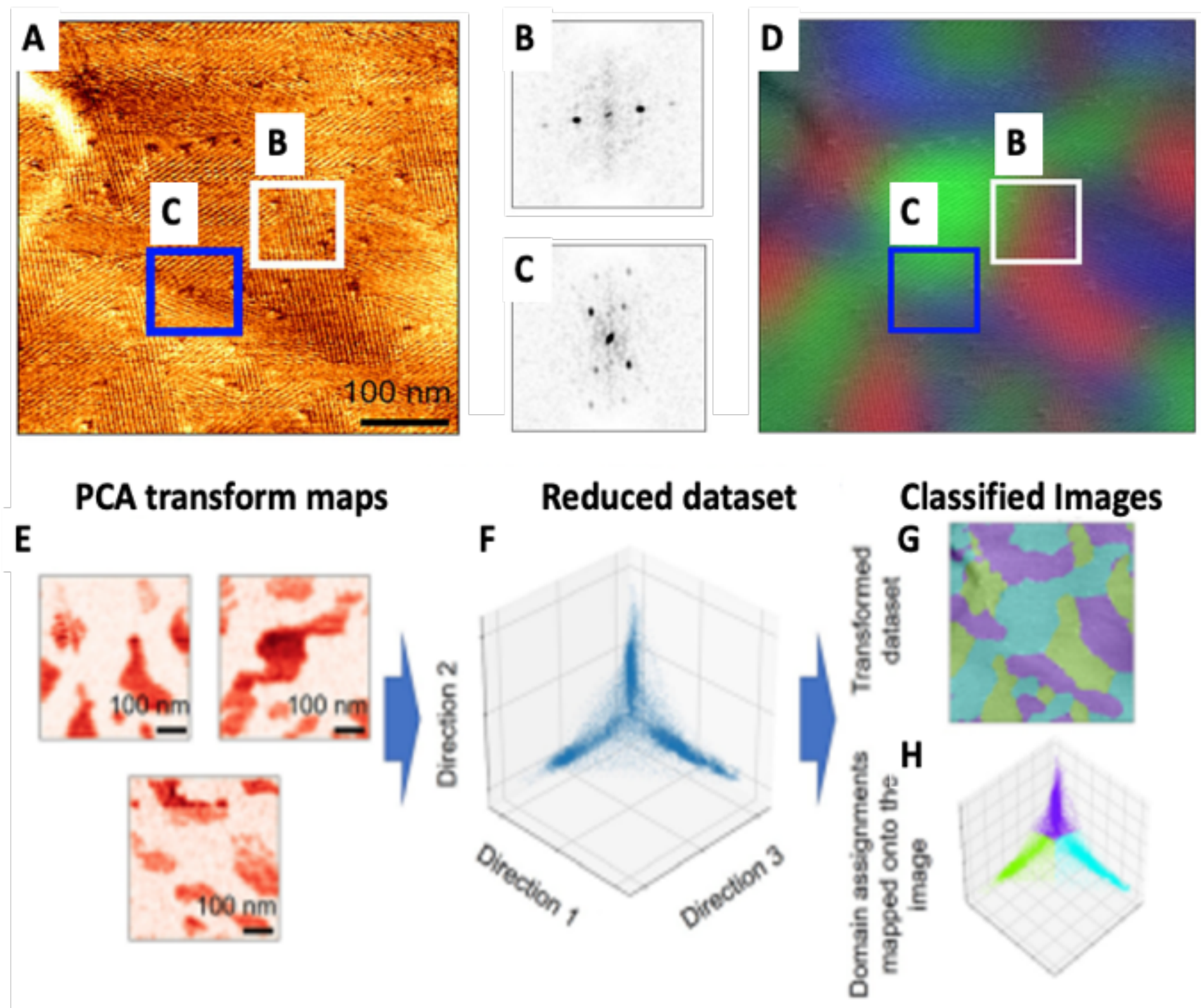

**Figure 2**: A) multidomain AFM image of EMI-TFSI ionic liquid self-assembled on a graphite surface with preferential crystallographic orientation. (B, C) structural descriptors based on the absolute value of the Fast Fourier transform (FFT) of the windowed image is used for classification. D) The output descriptor map overlaid on the AFM image, where red, green, and blue indicate domain orientations. E) PCA-transformed maps, F) feature space, G) original map is colored according to the, (H) segmentation of the feature space— in this case, it was done by the coordinate of the point. (Reproduced with pending permission from Borodinov et al[156]).

This workflow, applied to AFM images of ionic liquid layers on graphite and on boron nitride, demonstrates flexibility by allowing adjustments to different datasets and noise levels (**Figure 2**). The workflow accommodates expert insights into material organization by introducing transforms highlighting specific features, aligning with the principles of supervised machine learning. As another example by Belianinov et al[157], big data workflows have been developed to handle automated data collection, transfer, and analysis on high-performance computing (HPC) systems. These workflows involve converting raw microscopy data, into HPC-suitable formats and applying statistical methods, including PCA, k-means clustering, and Bayesian de-mixing, for dimensionality reduction and pattern recognition.  A third example by Valleti et al[158] involves workflows used to analyze atomically resolved scanning transmission electron microscopy images. Sub-images centered on specific atomic units, rather than an ideal lattice, were selected as local descriptors to account for material and microscope distortions. Rotationally invariant Variational Autoencoders (rVAEs) were employed to analyze ferroic materials under non-ideal imaging conditions, with latent spaces encoding rotation, offsets, and distortions. rVAE architectures featured symmetric encoder-decoder pairs trained to minimize reconstruction error and Kullback–Leibler[159] divergence. Clustering methods, such as k-means and Gaussian mixture modeling, were applied to segment sub-populations in the data. Dimensionality reduction techniques, including PCA, t-SNE, MDS, and ISO, were used to visualize sub-images in latent spaces. A Deep Convolutional Neural Network with a U-Net architecture was used for semantic segmentation, identifying atomic sites based on local contrast. Finally, latent variable maps were analyzed to visualize B-site cations in latent spaces. These workflows were applied to HAADF-STEM (High-Angle Annular Dark-Field Scanning Transmission Electron Microscopy) images of rhombohedral $BiFeO_3$ (BFO) thin films and BFO–LSMO superlattices, providing insights into structural and chemical variations.[158, 160, 161]

In these examples and universally across the field, the image analysis towards extraction of materials-specific insights represents a set of disparate operations, each with corresponding hyperparameters, that are suggested and optimized by human. Overall, these are extremely time consuming, prone to operator biases, and as such do not allow easy benchmarking, comparison, or implementation in the real-time image analytics to assist human decision-making. Hence, the key challenge is the automated construction of the workflows via sequential decision-making.

## IV. Reward driven workflows

The construction of the image analysis workflows and their parameter optimization can be represented as a decision-making process in the space of possible image analysis steps. Correspondingly, transition from human-driven to automated workflow construction requires solution of two problems. The first is the parameter optimization of the defined workflows, i.e. image operation sequence. The second is construction of the workflows from the set of image analysis operations. We pose that both problems can be fully automated if the reward function(s) driving image analysis can be defined. It enables the creation of explainable and entirely unbiased workflows while also facilitating the mapping of standard decision-making approaches. In this section, we define the general concept of the reward driven workflows and provide several examples illustrating all aspects of this concept.

### IV.1. General concept: reward

We pose that the result of the image analysis can be evaluated via the reward function, currently implicitly used by the human operators to guide the process. For the automated workflow building, this implicit reward function must be represented as an explicit function of the analysis results. Here, we summarize the current state-of-the-art in reward functions from ML perspective.

A reward function is a key concept in artificial intelligence and optimization, defining a measure of success for an agent's actions, allowing it to understand which behaviors are desirable. Essentially, the reward function assigns a value to different states or transitions within an environment, guiding the agent towards maximizing cumulative rewards over time. This approach aligns with the goal of achieving optimal behavior, as the agent learns which actions lead to favorable outcomes. In practical terms, a well-designed reward function can effectively guide an analysis or decision-making process towards success.

One notable extension of this concept is reward shaping, where additional rewards, or pseudo rewards, are introduced to facilitate learning. Reward shaping is used to modify the immediate reward in a way that makes it more indicative of the desired actions, which helps an agent learn complex behaviors more efficiently. This technique can accelerate the learning process without altering the final optimal policy, ensuring that the goal remains unchanged while leading the agent along a more effective learning path. The shaping theorem supports this concept by allowing a potential-based term to be added to the reward function, preserving the optimal policy while aiding in better decision-making.[162]

Building on these principles, reward-driven workflows represent a new paradigm for image analysis by utilizing reward functions to optimize the entire analysis process. In this framework, the reward function acts as a quantitative measure of success, allowing the workflow to be optimized in an adaptive manner. Unlike traditional methods that rely on fixed procedures or require expert-driven manual parameter tuning, this approach enables the dynamic adjustment of operations and hyperparameters. Through stochastic optimization frameworks like Bayesian Optimization[163], Monte Carlo decision trees[164], or reinforcement learning[165], reward-driven workflows can guide the iterative myopic or complex, multi-stage analyses, making them robust, flexible, and explainable.

The adaptability of reward-driven workflows makes them well-suited for real-time applications, where rapid adjustments are needed based on evolving data. By framing analysis as an optimization problem, they combine the strengths of classical and modern approaches. They automate while preserving interpretability, providing a clear, goal-driven structure for image analysis. This balance makes them powerful for tasks ranging from post-acquisition processing to real-time imaging, bridging gaps left by supervised and unsupervised methods.

### IV.2. Rewards in historical perspective

**Table 2**: partial table of past papers, accomplishment, including the concept of rewards.

| ACCOMPLISHMENT | REWARDS | REFERENCES |
|---|---|---|
| **MICROSCOPE OPTIMIZATION** | **Image quality** represented by image contrast, noise level, clarity of certain atomic or molecular features, Aberration correction. | Ref. [166-175] |
| **OBJECT DETECTION** | **Detecting** the presence and precise location of specific atomic and molecular features. | Ref. [38, 50, 61, 62, 64, 176-181] |
| **ATOM MANIPULATION** | **Atomic fabrication:** Movement of specific atomic and molecular features. | Ref. [37, 182-187] |
| **PHYSICS DISCOVERY** | **Structure-property relationship:** Physical properties extracted from imaging or spectral measurements. | Ref. [60, 188-193] |

All previous work in automating microscopy with machine learning—whether explicitly or implicitly relies on the concept of rewards (see **Table 2** for an overview). For instance, in the automated optimization of optical microscopes,[170] rewards such as image contrast and noise level have been used to achieve optimal imaging conditions.

Similarly, automated aberration correction in electron microscopes and scanning optimization/tip conditioning in scanning probe microscopes employ reward signals to guide adjustments (e.g., minimizing noise or maximizing feature clarity). [166-169, 171, 172] Another recurring example is the detection of specific atomic or molecular features, where convolution scores and other structural metrics act as implicit rewards to drive segmentation or localization workflows. [38, 50, 61, 62, 64, 176, 177] These ideas extend to tasks that are challenging for human operators—like moving individual atoms or molecules in scanning probe microscopy—where reward functions reflect successful positioning or assembly of desired structures.[37, 182] At higher levels, rewards extracted from images or spectroscopy measurements (e.g., EELS/EDX) facilitate automated discovery and optimization of new materials, linking observed data to meaningful physical properties.[60, 188, 189,

[173-175] Finally, atomic manipulation techniques further use reward feedback to carefully position atoms and molecules, supporting precise fabrication at the atomic scale.[183-187]

To illustrate how the general principles described in Section IV operate in practice, Section V presents a set of focused examples that demonstrate different aspects of reward-based workflow optimization within the level of automation discussed in Section I. These examples were selected to highlight the variety of ways in which reward functions can be defined, how parameter spaces can be explored, and how the resulting workflows adapt to diverse imaging scenarios. Each example serves as a concrete demonstration of the broader framework, showing how reward-driven methods translate abstract concepts into practical, automated image analysis strategies.

In all examples presented in this section, multi-objective optimization was performed using a genetic algorithm with a consistent configuration to enable fair comparison across workflows. Specifically, each optimization was initialized with a population of 100 candidate solutions and evolved over 10 generations, with reward evaluations performed at each generation to construct the corresponding Pareto fronts. This setting was found to be sufficient for convergence in all cases reported here, and the same optimization protocol was used across all examples unless otherwise noted.

## V. Reward-based optimization of a single workflow

The key step in building the reward-driven workflow is defining the reward functions. This can be built based on priori physical knowledge, human heuristics, crowdsourcing, and combinations thereof, and is highly domain specific. We note that the number of scenarios for which reward functions can be built is more limited than for supervised learning, and as such this approach is complementary to data-driven supervised learning. The reward workflow can also be used as a wrapper around the pretrained models, such as Segment Anything Model from Meta.[86] Workflow design can be built on multiple reward functions. The second aspect is defining the parameter space of the workflow, that can be represented as direct product of parameter space of all workflow steps. Below, we demonstrate several examples of reward-based workflows.

### V.1. Example 1: Reward-Optimized LoG for Atomic Positioning (LoG*)

The main challenge in this example is optimizing image analysis workflows for atom detection in microscopy, which typically either rely on manual tuning by experts or require extensive training and labeled data for DCNN models. Instead of relying on fixed procedures or expert-driven tuning, reward-driven workflows use well-defined reward functions as measures of success to guide the optimization process.[194]

The reward function quantitatively evaluates how well the analysis meets predefined objectives, in this case accurate atom detection. By iteratively optimizing these objectives using global optimization strategies such as Bayesian optimization or genetic algorithms, the workflow adaptively explores the parameter space and updates processing steps to improve performance. For this purpose, optimization of the conventional Laplacian of Gaussian (LoG) algorithm hyperparameters for atom finding were studied as a model. This approach is characterized by a set of control parameters including length scales $\sigma_{min}$ and $\sigma_{max}$, threshold ($T$), and overlap ($\theta$), which define its parameter space.

**Table 3:** Reward Function Objectives

| OBJECTIVE | DEFINITION | PURPOSE | OPTIMIZATION IMPACT |
|---|---|---|---|
| **QUALITY COUNT** | **Oracle A:** Expected atom count from physical knowledge (stoichiometry, lattice structure).<br>**Oracle B:** Expected atom count from a DCNN detector. | Ensures the detected number of atoms matches a trusted reference | Penalizes under- or over-counting and pushes the optimizer toward parameter sets that produce the correct number of atoms. |
| **ERROR** | Deviation of detected atom positions from the expected lattice geometry. | Ensures physically meaningful placement consistent with lattice spacing and symmetry and suppresses false positives. | Penalizes lattice-inconsistent detections and guides the optimizer toward spatially accurate results |

As summarized in **Table 3**, the reward function consists of two objectives. The first, *objective_1* or Quality Count, measures how well the number of detected atoms matches the expected count defined by the Oracle, which is based on physics-derived blob counts. The second, *objective_2* or Error, evaluates how closely the detected atoms follow the physical lattice structure. The optimized version of the LoG atom-detection algorithm, referred to as LoG*, is formulated as a multi-objective optimization problem that seeks to jointly minimize both *objective_1* and *objective_2* across the image processing parameter space defined by $\sigma_{min}$, $\sigma_{max}$, $T$, $\theta$. To benchmark accuracy, we use a reference standard known as the *Oracle*. Oracle_A corresponds to the physics-based atom count, while Oracle_B uses the atom count generated by the DCNN detector. As illustrated in **Figure 3**, a set of optimal solutions was obtained, where improving one objective comes at the cost of degrading the other. This framework establishes a delicate balance between the two competing objectives, leading to the identification of an optimal hyperparameter set for the LoG function.

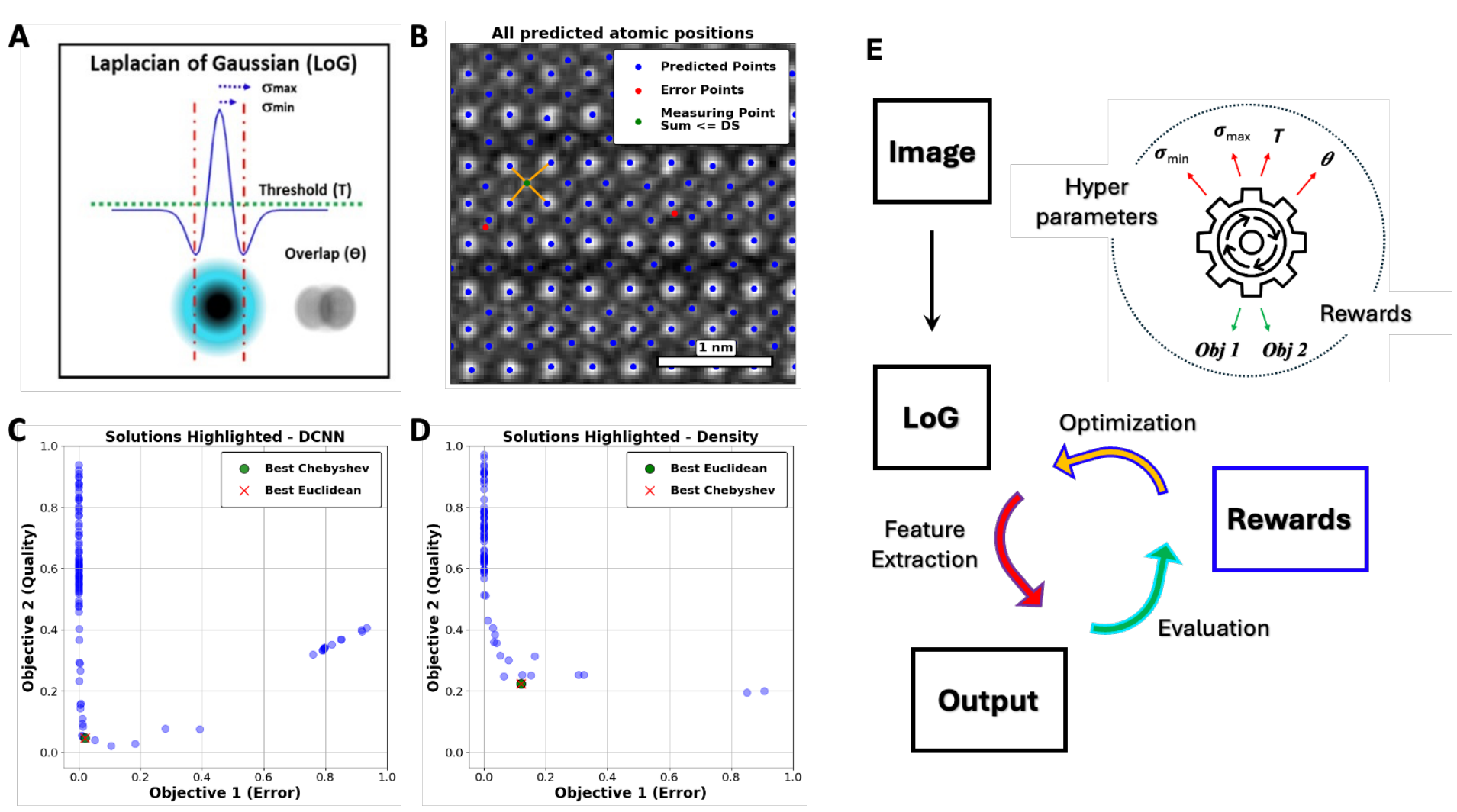

**Figure 3:** A) Laplacian of Gaussian hyper-parameters, B) Detected atoms and their nearest neighbor connections. Atoms marked in red indicate those with a sum of distances to their four nearest neighbors less than actual lattice length, thus flagged as errors, C) Pareto Frontier solutions with respect to *Oracle_A*, and D) Pareto Frontier solutions with respect to *Oracle_B*). E) LoG* optimization workflow. (Reproduced with pending permission from Barakati et al.)[194]

### V.2. Amorphous Regions: Reward-Guided Clustering and Identification

The second example illustrates identifying amorphous regions within an HAADF image caused by ion irradiation.[194] The traditional approach involves clustering detected atoms to identify variations or deviations in the lattice structure, typically using GMMs or similar clustering methods. However, this process relies on several hyperparameters, such as descriptor size for analysis and GMM-specific parameters like the covariance type, which significantly affect the results. The difficulty lies in determining whether the parameters were optimally set, as their effects only become evident at the end of the workflow, making it uncertain if the final output is truly accurate.

In the reward-driven workflow approach, we define physically meaningful reward functions for this analysis, allowing the rewards themselves to guide the optimization of parameters throughout the workflow. Two key objectives: *objective_1* (Compactness) and *objective_2* (Perimeter) of the clustered region were defined as two targets of the workflow. Compactness aims to create tightly bounded clusters, while minimizing regions perimeter ensures that structural deviations are localized. The parameter space includes the descriptor size height and width ($w_h$, $w_w$), which defines the spatial analysis range, and the *covariance_type* for the GMM, affecting how variance is represented. As shown in **Figure 4,** a set of optimal solutions was identified, highlighting a balance between objectives. Using a metric to select the best outcomes along the Pareto Frontier, the analysis effectively determined the most suitable descriptor size and covariance type for GMM clustering. This approach successfully mapped and identified areas within the material Yttrium Barium Copper Oxide (YBCO) substrate that show a higher likelihood of atomic deviation from predicted positions.

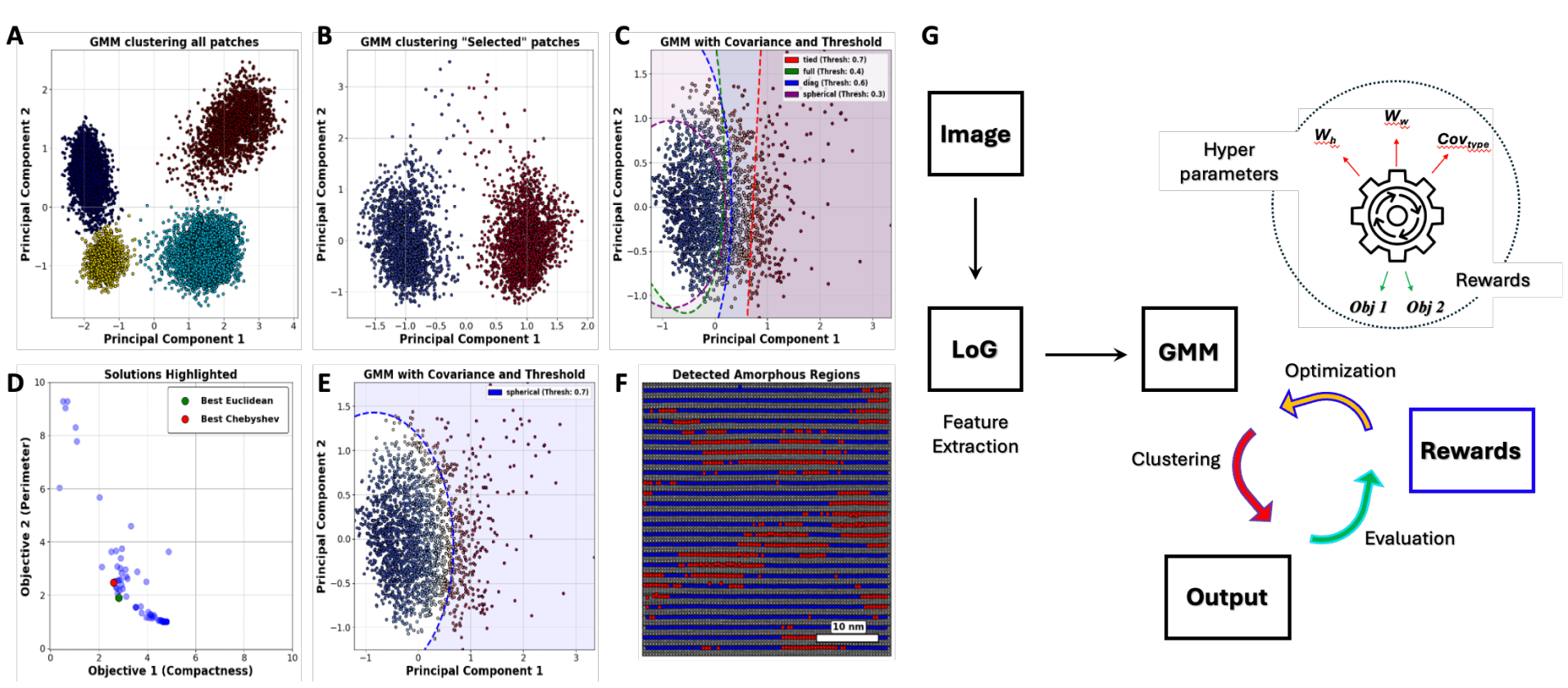

**Figure 4:** A) GMM clusters based on all the patches, providing 4 clusters with respect to 4 types of strong atoms in the YBCO structure. B) GMM clusters based on the patches centered on Ba atoms, presenting two types of Ba in the YBCO structure, C) GMM clusters based on only one type of Ba atoms, introducing some variety, which it can be differentiated by different values of threshold and covariance type in GMM clustering, D) Pareto Frontier solutions with respect to reward possession), E) Optimal threshold and covariance type achieved by multi-objective optimization (MOO) for GMM clustering, and F) Uncovered amorphous areas in the substrate, G) Amorphous regions detection workflow. (Reproduced with pending permission from Barakati et al.)[194]

### V.3. Ferroelectric Domains: Multi-Objective Reward Optimization for Domain Mapping

The third example illustrates identifying phases and ferroic domain walls in ferroelectric materials.[195] These walls represent the boundaries between different regions, or domains, where the polarization is oriented in different directions. The movement, interaction, and configuration of these domain walls directly influence the material's overall ferroelectric, piezoelectric, and dielectric properties, which are important for various technological applications, including memory devices, sensors, and actuators.

Conventional image analysis methods require extensive manual intervention to identify domain boundaries because these boundaries often lack strong contrast, are obscured by local distortions, and take on complex shapes. Moreover, the process involves many key hyperparameters such as filter settings, window sizes, and clustering parameters that greatly influence the results. These challenges make it difficult to achieve consistent and accurate identification of domain boundaries without constant fine-tuning by experts. **Figure 5** illustrates the impact of two hyperparameters: window size and the number of GMM components.

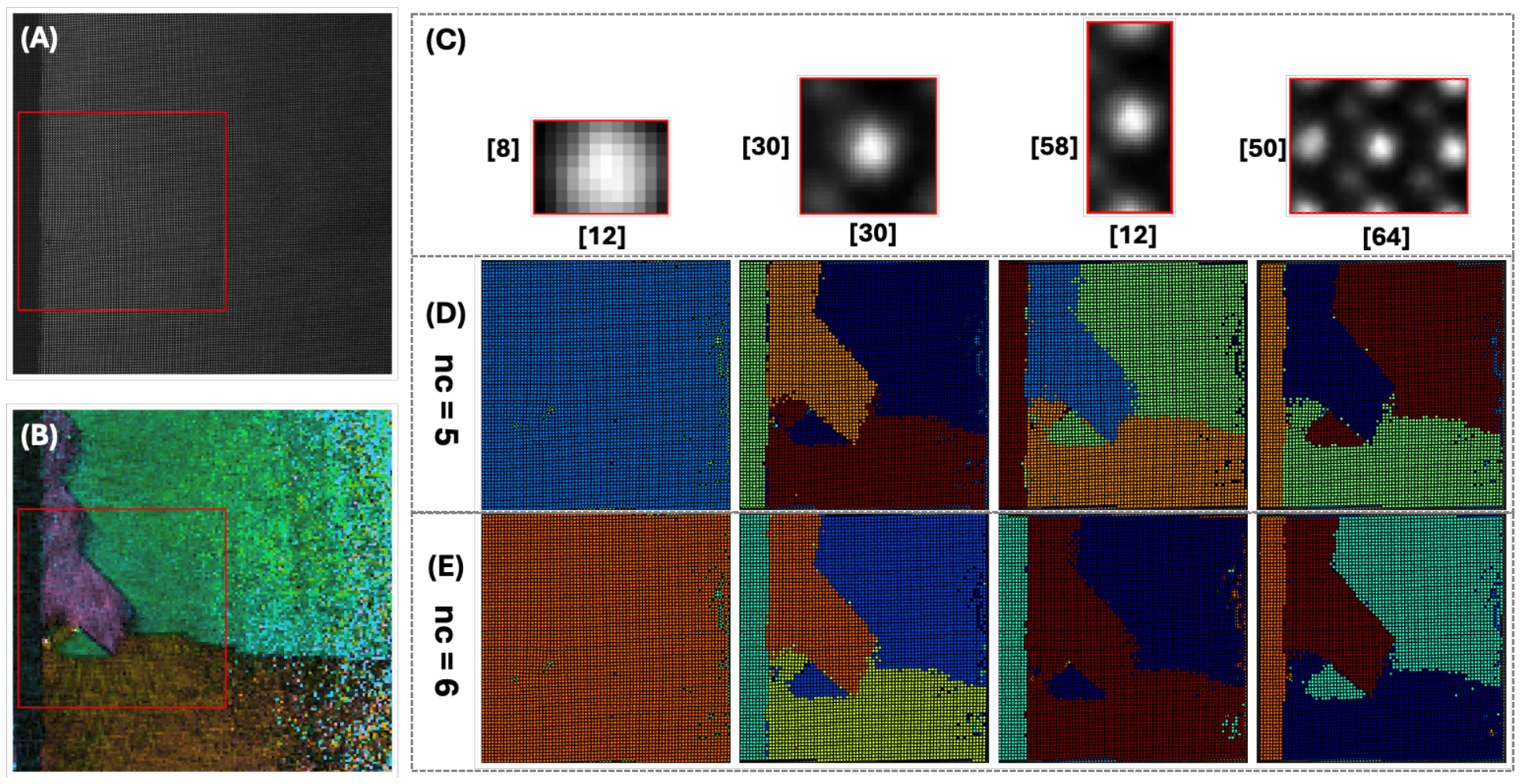

**Figure 5**: (A) HAADF image showing the selected region of interest used for analysis. (B) Ground truth polarization map of the region, illustrating the actual polarization distribution. (C) Selected descriptor examples. (D) Clustered regions using GMM with 5 fixed components, and (E) clustered regions using GMM with 6 fixed components, both illustrating the effects of window size selection and cluster count on segmentation accuracy and granularity. (Reproduced with pending permission from Barakati et al.)[195]

The parameter space for this workflow consists of three primary hyperparameters: window size dimensions ($w_h$, $w_w$), which define the size of the image patches centered on atomic columns, and the covariance type for the GMM clustering. Two reward functions were defined to guide the optimization within this parameter space. The first *objective_1* (Straightness) is aimed at minimizing the curvature of detected domain walls, encouraging the formation of smooth, continuous boundaries that are characteristic of stable ferroelectric domain walls. The second *objective_2* (Length) focuses on maximizing the continuity of detected domain walls, avoiding fragmentation, and ensuring a more accurate representation of extended structures. Together, these rewards create an optimization objective that balances the natural morphological characteristics of ferroelectric domain walls. Results showed in **Figure 6** implies that the reward-driven workflow successfully identified the optimal parameter combinations that provided the best trade-off between smoothness and continuity of domain walls.

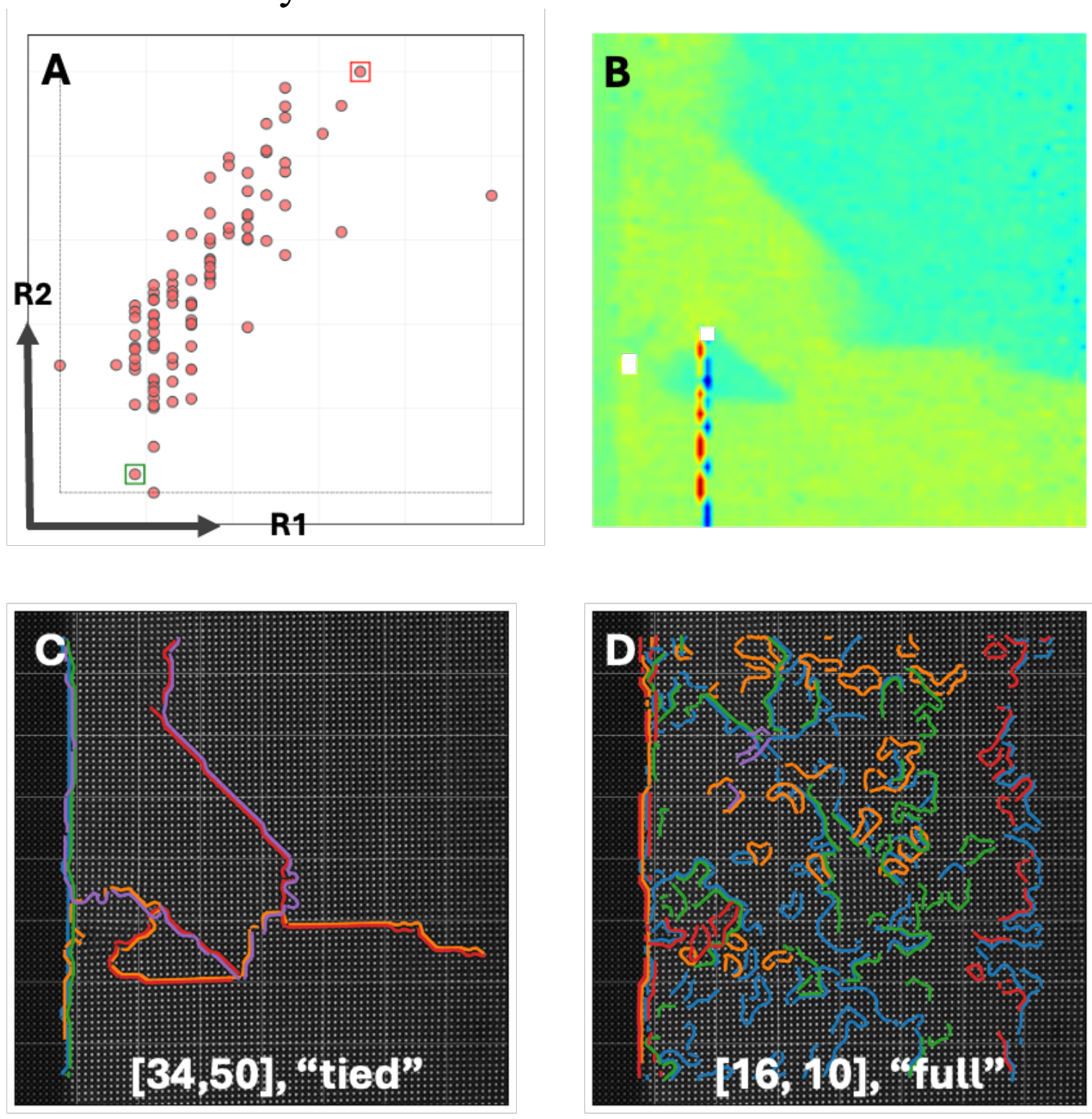

**Figure 6**: A) Pareto front solutions representing the trade-off between two rewards, $R_1$ (Straightness) and $R_2$ (Length). B) Ground truth polarization map. The red and green markers highlight the solutions selected by operator during analysis according to be prerequisites of the experiment, (C) Solution acquired by the workflow, showing the best possible trade-off between objectives of the experiment, (D) Another option in pareto front solution that operator has selected to explore the materials properties. (Reproduced with pending permission from Barakati et al.)[195]

### V.4. SAM*: Reward-Driven Fine-Tuning of the Segment Anything Model

This example addresses a key limitation of foundational models, which involve numerous non-transparent tuning parameters requiring extensive manual optimization, thereby restricting their applicability. To mitigate this issue, we apply a reward function-based approach to fine-tune the Meta SAM (Segment Anything Model)[86] framework, enhancing its adaptability for image segmentation in microscopy.

SAM includes over 10 key hyperparameters, such as "points_per_side", "pred_iou_thresh", "stability_score_thresh", and "box_nms_thresh", which govern segmentation precision, stability, and filtering. While these parameters provide flexibility, manually tuning them for domain-specific applications remains a challenge. Detailed descriptions of these parameters and their roles within SAM are provided in Ref.[196] (Barakati *et al.*). However, by leveraging the reward-driven workflow, we can systematically refine these hyperparameters and steer the model toward our specific objectives. Through well-defined reward functions, SAM can be dynamically optimized to prioritize desired segmentation characteristics, such as distinguishing between small and large particles, enhancing stability, or improving mask quality, thereby eliminating the need for manual parameter selection, and enabling task-specific adaptability.[196]

In this example, we aimed to optimize SAM on an AFM topography image of combinatorial Au-Co, by tuning the hyperparameters "points_per_side", "pred_iou_thresh", "stability_score_thresh" to capture features of varying sizes, guided by two reward functions. As presented in **Figure 7**, segmentation process has been fine-tuned to identify either small or large features. In this case, the key outcome is the optimized set of hyperparameters at each extreme—prioritizing *objective_1* leads to detecting only small particles, while maximizing *objective_2* results in identifying only large particles.

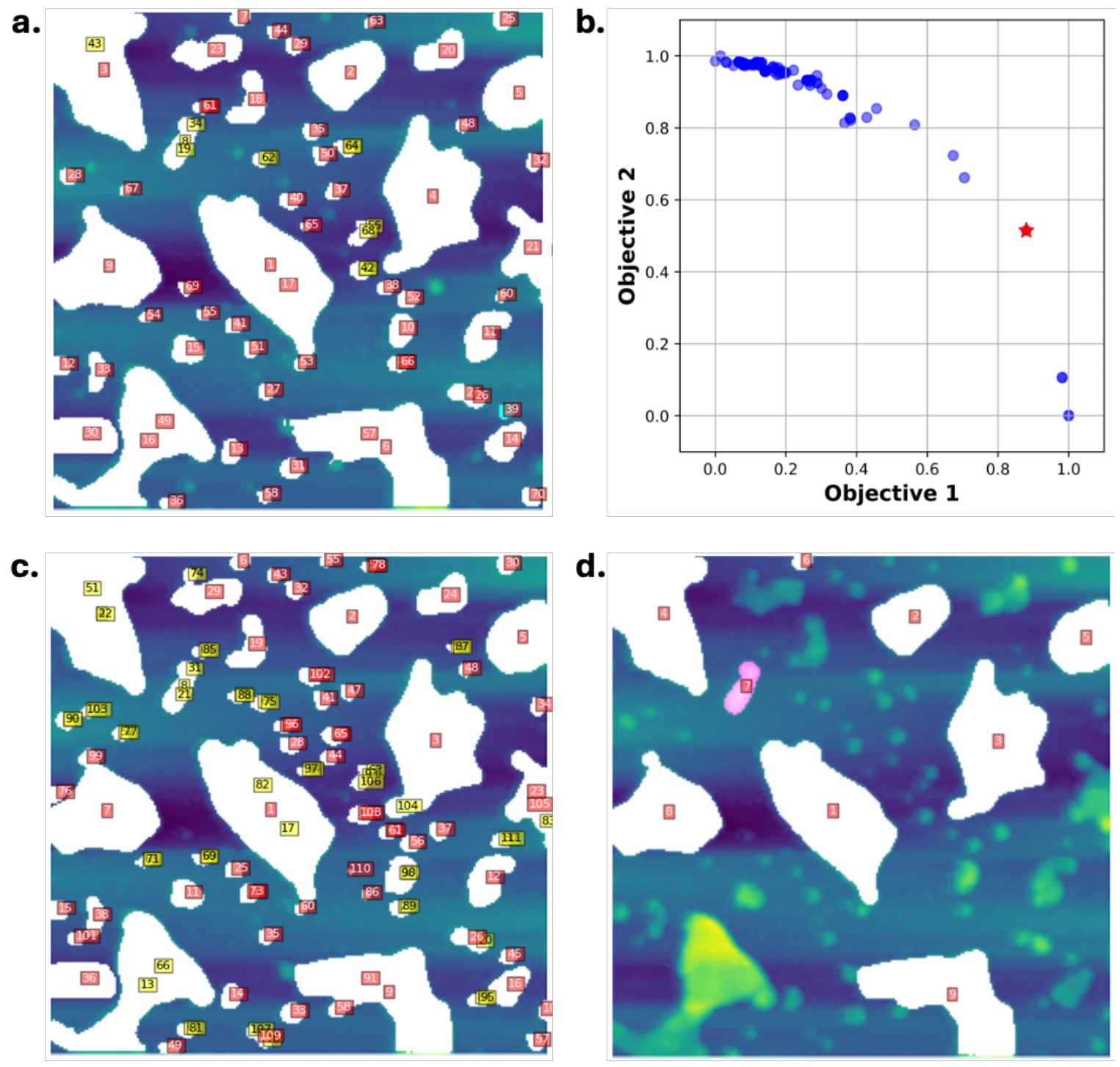

**Figure 7**: Segmented AFM topography image of Au-Co combinatorial library, a) Standard SAM segmentation result. b) Pareto front solutions obtained by adapted reward-driven workflow, balancing two objectives, c) Segmentation result favoring *objective_1*, prioritizing one aspect of the optimization, d) Segmentation result favoring *objective_2*, emphasizing the alternative optimization criterion.

The reward-based formulation provides a principled pathway for translating experimental intent into an optimization problem, thereby replacing manual hyperparameter selection with an automated, objective-driven search of the workflow parameter space. By explicitly encoding physical constraints, structural expectations, and task-specific priorities into reward functions, analytical pipelines are guided toward solutions that are consistent with underlying scientific knowledge rather than algorithm-dependent heuristics. This approach supports multi-objective trade-offs in a transparent manner and remains agnostic to the choice of analytical method, allowing both traditional algorithms and pretrained foundation models to be incorporated within the same optimization framework and directed toward physically meaningful experimental outcomes.

In all examples presented in this section, multi-objective optimization was carried out using a genetic algorithm with a consistent configuration to enable fair comparison across workflows. Each optimization was initialized with a population of 100 candidate solutions and evolved over 10 generations, with reward evaluations performed at each generation to construct the corresponding Pareto fronts. This setting was sufficient to reach convergence in all cases reported here, and the same optimization protocol was applied across all examples unless otherwise noted.

### V.5. Future opportunities

The integration of real time reward-driven image analysis into experimental workflows will not only open many new opportunities but also have the potential to change how experiments are conducted completely. Conventionally, researchers wait for the image scan or grid-map to finish before identifying interesting features, which is both time-consuming and lacks the flexibility of real-time experiment. With the real time image analysis integrated, operators will be able to identify features of interest in the real time and thus have the flexibility to take time-consuming measurement only at these locations. Moreover, it enables building advanced automated decision-making process for autonomous experimentation. For example, human operators can define a policy to automatically adjust the spectral parameters, including spectral time and resolution, number of measurements, magnitude of voltage pulse, according to the types of structural features detected by real time analysis.

Ideally, such framework should allow development of adaptive workflows that evolve dynamically with experimental progress, for example to refine their measurement strategies based on incoming data, continuously improving their accuracy and relevance. The ability to detect and analyze emergent phenomena, such as transient states or rare defects, in real-time could transform fields like materials science and biological imaging and pave the way for experiments that are not only more efficient but also more exploratory, enabling the discovery of phenomena that might otherwise be overlooked in static, pre-defined workflows. This is particularly critical for the fields with temporal constraints, such as tracking dynamic biological processes or monitoring time-sensitive chemical reactions. Below, we discuss possible strategies for the dynamic reward-based workflow construction mapping this process on decision-making in the space of possible classical and NN-based analysis operations.

## VI. Reward-based workflow constructions

The examples presented in sections **V.1-5** are the few-step workflows with the sequence of operations defined by human operator optimized in the joint parameter space of all functions via the reward function approach. Here, we discuss approaches for the construction of arbitrary length workflows, i.e. defining the set of image analysis operations with subsequent hyperparameter optimization, as exemplified in **Figure 8.** This is a classical sequential decision problem.[197] As such, we define the typical set of operations in workflow constructions and illustrate how these can be mapped on decision-making theory.

### VI.1. General workflow structure

The typical image analysis workflow proceeds through the set of sequential steps from global operations such as smoothing, subtraction, and contrast equalization to more complex keypoint detection, building descriptors, and descriptor analysis. In this process, the image is converted from the raw data set to the collection of representations with much richer semantic content. It is remarkable that this process is very similar to text analytics, starting from the optical character recognition tasks to the construction of words, part of speech tagging, etc., all the way to construction of the knowledge graphs.[198, 199]

The initial step typically differs from standard image processing techniques such as smoothing, background subtraction, and contrast equalization, which are applied globally to the image. Note that some of these steps are designed for human perception, while others can meaningfully affect the downstream image analysis operations.

The second step is identification of the keypoints. These can be (a) rectangular grid sampling/sliding window, defining dense keypoints such as atoms of an atomically resolved images, local maxima, structural keypoints identified by methods such as SIFT, ORB (Oriented FAST (Features from accelerated segment test) and Rotated BRIEF (Binary Robust Independent Elementary Features)), etc. keypoints can also be sparse, for example samples from grain boundaries and topological defects within the image. Keypoint definition is a strategy to incorporate human heuristic/reward in the analysis pipeline. Note that in the limiting case of rectangular grid, each pixel of the image can be interpreted as a keypoint.

With the keypoints defined, the next (optional) step is classification of the keypoints based on local signatures, for example intensity. Methods like SIFT or LoG create a descriptor vector before finding the keypoint, and hence keypoint classes can be formed based on these.

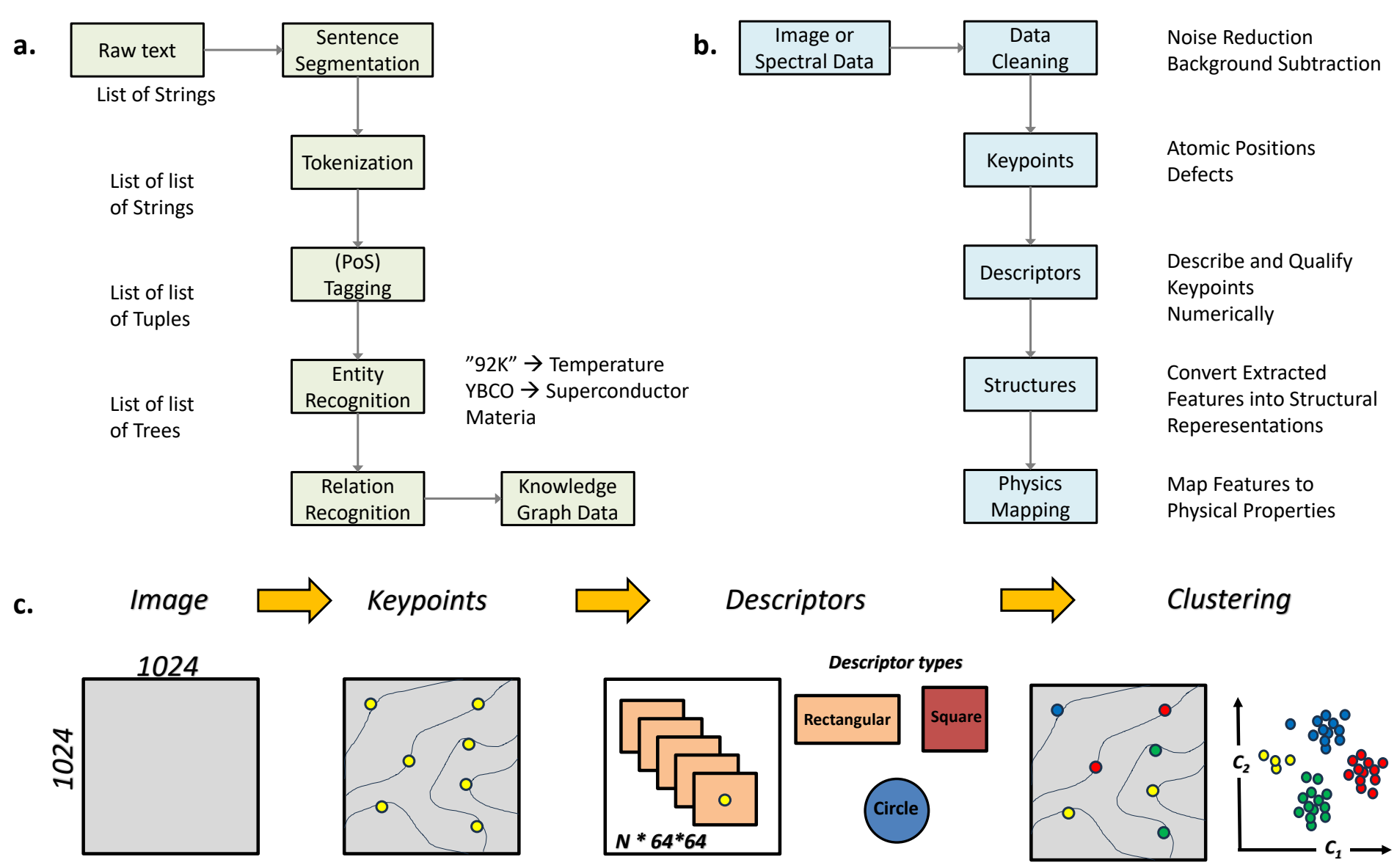

**Figure 8**: A comparative workflow representation: a) Natural Language Processing (NLP)[200] pipeline, illustrating key stages such as tokenization, embedding, and model training; b) Image Analysis pipeline, outlining steps from preprocessing and feature extraction to classification and segmentation, incorporating both classical and deep learning approaches, c) An example set of image analysis operations with subsequent hyperparameter optimization.

With the keypoints defined, the semantic meaning is added via the local descriptors. Simplest example of the descriptors is square image patches centered at the keypoints. However, these can be much more general and include rectangular image patches, rotated rectangular, circular, elliptical, and even non-continuous patches. Depending on the problem (e.g. time-lapsed video data) the descriptors can be defined at the video frame shifted compared the keypoint one, allowing for analysis of time dependent phenomena. It is also important to note that at this stage the analysis of images and more complex spectroscopic data sets becomes isomorphic. For example, spectroscopic data sets can be interpreted as keypoints corresponding to the dense measurement grids, and spectral data being the local descriptor. Hence subsequent discussion applies equally to the analysis of imaging and hyperspectral data.

With the descriptors formed, they can be transformed via a set of physics-based and data-based transforms. For example, image patches can be transformed via FFT, Hough, Radon transforms, Gabor filter banks, and other operations. The spectral and imaging data can be transformed via the physic-based model. Finally, the descriptor can be left invariant (identity transform). Following the transformations, the descriptors can be clustered, dimensionality reduced via linear or non-linear methods or classified. Many of these operations can be done sequentially. For example, the analysis workflow in **V.2 Amorphous regions** is based on a sequence of the clustering processes, whereas another example can be utilizing keypoint detection followed by feature extraction and clustering. In human built workflows the number of the analysis operations is generally limited; however, this limitation stems from the capability of human operator rather than represents objective limits on the workflow length.

The result of the generalized image analysis workflow is the transformation of the image into the set of 2D maps where each keypoint is associated with a (small) set of dimensionally reduced variables. For example, it can be a cluster label for each descriptor, set of the PCA components or the VAE latent variables, etc. Common for these is that by design, the resultant maps are 2D and their number is finite, making them amenable to human interpretation and subsequent decision-making. Note that in principle it can be sparse images (like properties along domain walls and topological defects); similarly, this approach can be extended to 3D images, such as tomography or tilt- or focal series reconstructions.[201-203]

For the workflows designed above, the number of possible operations at each analysis step can be significant, and the analysis results sensitively depend on each step. For example, the definition of the descriptors at the early stage of workflow will sensitively affect the downstream results. This renders workflow construction complicated, biased, and time consuming. Below, we discuss the desiderata and some possible strategies for workflow construction.

### VI.2. Data structure

In scientific imaging and analysis, well-defined data structures are essential for organizing and preserving both raw data and the associated workflows. Images typically require classical formats such as TIFF or JPEG for storage, while hierarchical data formats like HDF5 are crucial for maintaining metadata, ensuring reproducibility, and enabling efficient access to large datasets. Similarly, knowledge graphs are widely used for text-based representations, allowing structured storage and retrieval of relationships between entities. However, existing formats primarily focus on static data storage and lack a dedicated structure for representing the dynamic nature of image analysis workflows. We propose that reward-based image analysis requires constructing data structure specifically designed to capture and store workflow execution, making processes traceable while enabling iterative optimization and automation. By integrating workflow metadata, parameter history, and analysis steps into a structured format, this approach will facilitate reproducibility, adaptation, and improvement of experimental and computational pipelines, bridging the gap between data storage and real-time decision-making in scientific research.

The method contains three types of objects: State, Reward, Workflow components. We assume that reward function is available in the end of workflow. However, certain decision-making algorithms require heurists how close we are to this reward at intermediate workflow stages. These heuristic functions can then be used in schemes such as A* search or reinforcement learning, as will be discussed in section **VI.4.**

The state of the system at any given point in the workflow is encapsulated by the data structure [195], where: $\boldsymbol{K}$ (Keypoints): is a $[\kappa * 2]$ matrix representing the coordinates of the detected Keypoints within the image. $\boldsymbol{D}$ (Descriptors): A tuple [H, W] for rectangular descriptors (height and width) or [R] for circular descriptors (radius), capturing the geometric attributes of each descriptor. $\vec{\boldsymbol{c}}$ (Clusters): A vector with integer entries indicating the cluster assignment for each Keypoint, reflecting the grouping determined by the clustering algorithm. The state evolves as it moves along different stages of the workflow. At the end of the workflow, we finally arrive at the *optimal state* $\{\boldsymbol{K}, \boldsymbol{D}, \vec{\boldsymbol{c}}\}$, which maximizes the total reward.

The reward mechanism is pivotal for guiding the optimization process towards desirable workflow configurations. Rewards are computed based on the current state $\{\boldsymbol{K}, \boldsymbol{D}, \vec{\boldsymbol{c}}\}$, with a focus on physical properties and domain knowledge. Rewards can be used to optimize workflows. The overarching goal is to identify the optimal state $\{\boldsymbol{K}, \boldsymbol{D}, \vec{\boldsymbol{c}}\}$ that maximizes the reward. Achieving this involves (a) **Building workflow**, i.e. hoosing the appropriate algorithms for each workflow component (Keypoint finder, descriptor creator, dimensionality reduction, clustering) and configuring their parameters, and (b) **Parameter Tuning**: Adjusting the workflow parameters iteratively to refine the state towards optimality, guided by the reward signals.

Additionally, certain strategies for decision-making require roll outs functions that can be expected to approximate expected rewards. We discuss these in section VI.4. Finally, we note that workflow optimization may require the definition of certain operations. For example, clustering of the keypoints or descriptors is a non-differentiable operation, since changing upstream conditions can affect the assignment of the cluster label. Here, possible strategies may include the introduction of the anchor labels that have predefined classes, or majority-based indexing such that major class always have lowest label index.

### VI.3. Full Workflow Design

Here we discuss the methods based on the full workflow construction, with the subsequent optimization in the joint parameter space of all operations by given reward function.

**Combinatorial search**. For short workflows and relatively small number of possible operations, workflows can be form combinatorically to represent all possible sequences of sequentially compatible image analysis steps. Once formed, each of the workflows can be jointly optimized and results can be compared based on the reward values.

**Genetic algorithms**. The workflows can be defined via genetic algorithms. Here, several workflows are formed based on human inputs or randomly. The mutation and cross-over operations can be defined in a way to follow the desiderate. Mutation operations are relatively simple to define within similar operation classes (e.g. all clustering algorithms yield cluster label). The cross-over operations require definition that maintains the data continuity along the workflow. Note that Genetic Algorithm (GA)[197] can be combined with more complex optimization algorithms.

Each candidate solution (individual) in the GA population represents a unique set of workflow parameters. To effectively apply GA, we must encode these parameters into a chromosome structure. The chromosome is a concatenation of all workflow parameters, each encoded as genes. For the example workflow, the parameters to optimize are:

i. **KeyPoint Finder (Blob Optimizer) Parameters:** Min Sigma ($\sigma_{min}$), Max Sigma ($\sigma_{max}$), Threshold ($T$)
ii. **Descriptor Creator (Rectangle) Parameters:** Height ($H$), Width ($W$)
iii. **Dimensionality Reduction (PCA) Parameters:** Number of Principal Components ($PC$)
iv. **Clustering Algorithm (GMM) Parameters:** Number of Classes ($K$), Covariance Type (*cov_type*): Categorical

**Chromosome Representation** *Chromosome* = [$\sigma_{min}, \sigma_{max}, T, H, W, PC, K, cov_type$]

The fitness function evaluates each individual based on the workflow's performance, quantified by the reward metric. These settings provide a balance between exploration of the search space and exploitation of promising solutions.

### VI.4. Sequential workflow design

For sufficiently long workflows, the end-to-end construction can be intractable do to the exponentially large search spaces. In this case, the workflows can be constructed by engineering engineer heuristics that determines how close we are for the problem of workflow construction.

**Decision trees.** Workflow optimization can alternatively be achieved by modeling the process as a decision tree like algorithms including Monte Carlo Decision Trees (MCDT), A* search, etc. In this approach, the workflow for keypoint detection and clustering are structured as a series of decision points, where each node in the tree represents a specific operation selection followed by parameter optimization, such as values for $\sigma_{min}, \sigma_{max}$, or radius as shown in **Figure 9**. Each branch corresponds to a potential action, adjusting a parameter within a continuous space. At the leaf nodes, the performance of the workflow is evaluated based on defined metrics or rewards, such as measures of straightness or uniformity. This hierarchical structure allows for systematic exploration and optimization of parameters to enhance workflow performance.

Strategies for optimizing workflows require the roll-out functions for reward estimation and parameter tuning of incomplete tree (where reward function has not been attained). A roll-out function estimates the potential reward associated with taking a specific action and propagating it through the decision tree. By approximating rewards for downstream outcomes, these functions guide the policy by providing a heuristic for navigating the continuous space effectively. This approach bridges the gap between decision-making in discrete and continuous domains, enabling more robust optimization strategies in complex workflows.

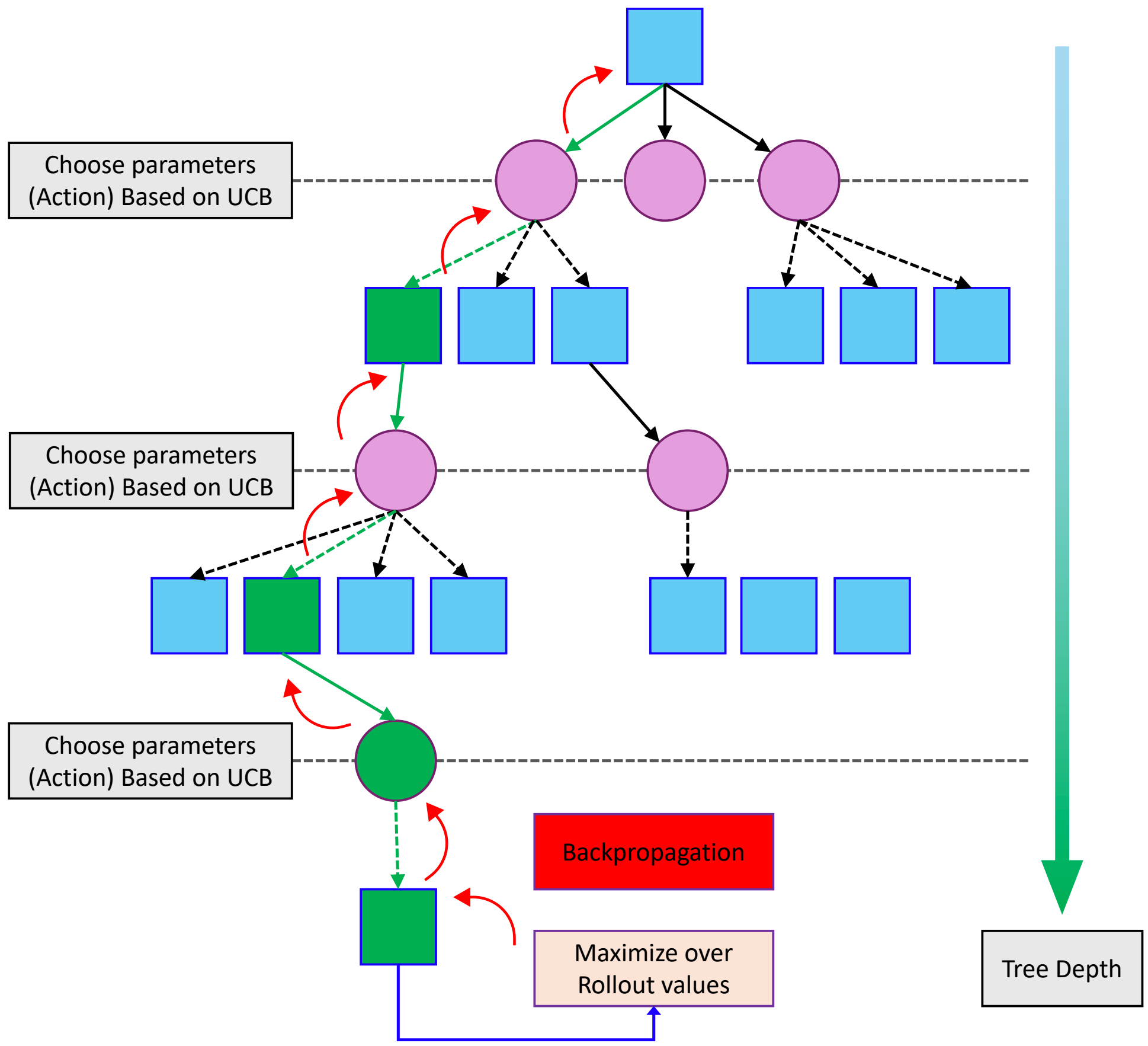

**Figure 9**: Markov Chain Decision Tree workflow. Square nodes represent states, while circular nodes denote decisions. A UCB-based exploration strategy is used to sample actions (decisions). The goal is to learn the parameters that maximize the final reward (rollout values).

While generally development of the roll-out functions can be a separate challenge, we pose that PCA can be used as a simple roll-out function with the argmax of the R(PCA) being a roll-out function.

**Reinforcement learning**. We pose that the roll-out functions can also be used as a reward function for the reinforcement learning type algorithms. As discussed in previous sections the rewards are achieved from the state $\{\boldsymbol{K}, \boldsymbol{D}, \vec{\boldsymbol{c}}\}$. This rewards until now is myopic and thus it is easy to fall into suboptimal workflow configuration. Reinforcement learning solves this problem by learning a policy (The parameters of the workflow) by optimizing over the total return. Where total return is defined by:

$$G_t = R_{t+1} + \gamma R_{t+2} + \gamma^2 R_{t+3} + \cdots = \sum_{k=0}^{\infty} \gamma^k R_{t+k+1} \quad (2)$$

Where: $G_t$ : Total return starting from time step 1, $R_{t+1} + \gamma R_{t+2} + \cdots$ : Rewards received at successive time steps, *and* $\gamma$ : is discount factor ($0 \leq \gamma \leq 1$) which determines importance of future rewards. And the reward function can be same roll-out functions.

## VII. Comparison of reward driven workflows vs. DCNNs

Benchmarking is a critical step for optimizing autonomous discovery systems and ensuring their robustness before real-world deployment. This involves systematically evaluating components like reward functions, seeding strategies, and workflow combinations, while measuring computational efficiency and scalability. Extensive validation using pre-acquired datasets is essential to establish baseline performance, assess workflow reliability, and ensure adaptability across modalities. Key benchmarks can include workflow efficiency, extraction of meaningful physics insights, exploration of trade-offs via Pareto fronts, cross-modality generalization, and the development of reliable performance metrics.

A detailed comparison of Reward-Driven Workflows (RDW) and DCNN is presented in **Table 4**, highlighting their strengths, limitations, and suitability for various tasks in real-time imaging and data analysis.

**Table 4**: Comparison of reward driven workflows vs. DCNNs

| Aspect | Reward-Driven Workflows | DCNNs |
|---|---|---|
| Real-Time Adaptation | Excellent for dynamic tasks with real-time feedback | Challenging due to computational demands |
| Ease of Interpretation | High, due to transparent reward metrics | Low, often a 'black box |
| Data Requirements | Low, works well with sparse datasets | High, requires extensive labeled data |
| Computational Demand | Low to moderate, depending on reward complexity | High, especially during training |
| Versatility | Limited to well-defined tasks. | High, with transferable features via pre-training |
| Scalability | Moderate, requires redefinition for new tasks | Excellent with sufficient data and compute |
| Setup Effort | Effort required for reward engineering | Significant for model training and hyperparameter tuning. |

Evaluating the model's detection performance under different noise levels helps assess its robustness and reliability. Successfully detecting atoms amidst significant Gaussian noise demonstrates the model's stability and resilience, ensuring reliable performance in practical, noisy microscopy environments. The LoG* method adapted to noise levels by adjusting hyperparameters, maintaining detection accuracy, while DCNN models produced false positives due to noise. As shown in **Figure 10**, the LoG* demonstrated greater stability and resistance to noise-induced misidentifications compared to DCNNs.

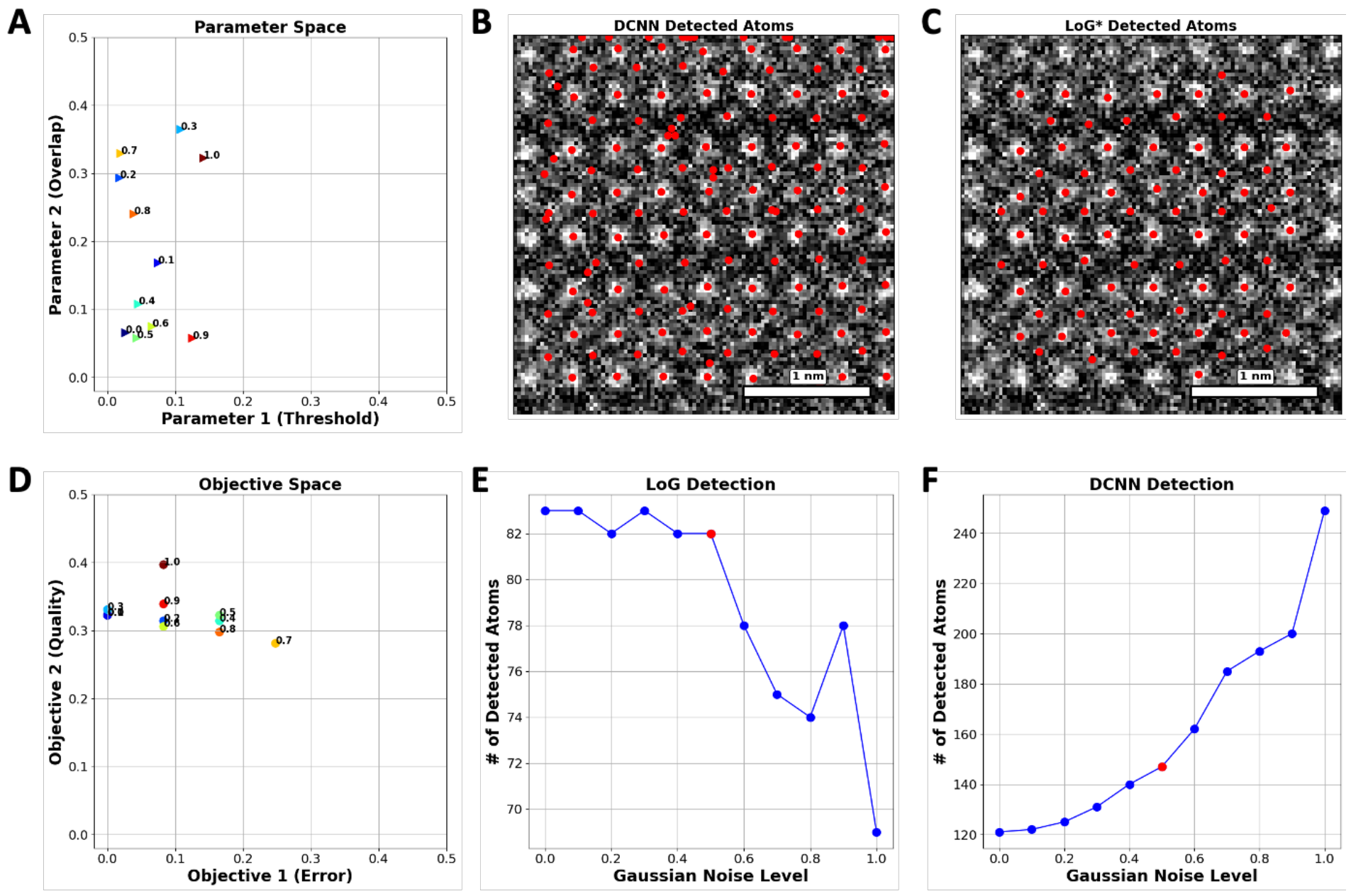

**Figure 10:** A) Optimal hyper-parameter space changes verses noise level in the LoG* optimized method, B) Detected atoms using the LoG* optimized method on the image with a moderate noise level, C) Detected atoms using a DCNN model scattered on the image, D) Optimal objective space change verses noise level in the LoG* optimized method, E) Number of detected atoms versus Gaussian noise level using the LoG* optimized method, F) Number of detected atoms versus Gaussian noise level using a DCNN model. (Reproduced with pending permission from Barakati et al.)[194]

## VIII. Reward generation and open sourcing

The concept of rewards is central to the progress and application of reinforcement learning and reward-driven workflows. It entails formulating a quantifiable function that assesses and provides feedback on an agent's performance or outcomes, ensuring alignment with predefined objectives. In scientific workflows, this process often involves defining metrics such as accuracy, efficiency, smoothness, or conformity to experimental goals. However, two challenges arise: reward functions are typically task-specific, requiring redefinition for new problems, and their design demands domain knowledge and iterative refinement to avoid unintended outcomes.

To address these, strategies include modular reward design by breaking rewards into reusable sub-functions, standardizing templates for common tasks, and open-sourcing libraries of pre-defined rewards to save time and promote collaboration. Automated reward tuning using Large Language Models (LLMs) can significantly reduce manual effort by generating, refining, and adapting reward functions based on user goals and observed outcomes. LLMs enable natural language inputs for intuitive reward design, making the process accessible even to non-experts. Additionally, community collaboration can be enhanced through open-source repositories, shared datasets, interactive platforms for reward function exchange, and virtual tools powered by LLMs that facilitate co-creation and iteration of reward structures across diverse applications.

### VIII.1. Physics based rewards.

This approach is particularly effective in domains governed by well-defined physical laws, such as materials science, microscopy, or fluid dynamics. Rewards are defined using measurable parameters like energy, force, or system stability and incorporate equations or models representing desired outcomes. For example, in microscopy, a reward function might aim to maximize contrast or minimize image noise by penalizing deviations from expected physical constraints. While this method ensures scientifically valid outcomes and reduces reliance on large datasets, it requires deep domain expertise to define accurate metrics and may not generalize easily to tasks without clear physical models.

We note that the problem of optimal data representation for human perception is intrinsically linked to the challenge of identifying the best underlying physical descriptors. The reason for this is that the desiderata for effective human perception, such as parsimony, interpretability, and generalizability, align closely with the criteria for robust physical models. In fact, the development of physical theories is often driven by these very criteria, as they enable the abstraction and simplification of complex systems into comprehensible and predictive frameworks. For example, over the past several years, there has been growing interest in employing non-linear dimensionality reduction methods, such as VAEs[204-209], to extract meaningful features from these complex systems.[194, 195, 210, 211] However, the VAE based representations are often non-unique, and require multiple hyperparameters. In this case, the reward function(s) can define the heuristic and physical constraints on final analysis results. These for example can include the total number of atoms, straightness of identified image segments (i.e. topological defects), histogram of identified regions, etc. As such, they can directly embody the constraints such as interfacial free energy minimization, Gibbs phase rule, bond-valence constraints, Pauling rules, etc. Once the reward functions are defined, image or spectral analysis becomes an optimization problem of construction of the image analysis workflow in the combinatorial space of sequential operations, and optimization of the corresponding hyperparameters. In practice, convergence and stability can be affected by several factors, most importantly the way rewards are defined and balanced during optimization. Certain failure modes related to noise have also been explored in earlier studies.[190]

### VIII.2. Crowdsourcing and human heuristics

This approach taps into collective human expertise to define and refine reward functions, particularly in tasks where physical models are incomplete or subjective insights are necessary. This approach involves gathering feedback from experts or laypersons to score task outcomes or develop heuristics that guide optimization. For instance, in image segmentation, human evaluators might assess outputs based on perceived clarity or accuracy, and this input can fine-tune reward functions. The method captures nuanced, subjective criteria that are difficult to quantify mathematically, fostering adaptability. However, it is time-intensive and may yield inconsistent results due to variability in human judgment.

### VIII.3. Model systems and inverse RL

This approach leverages simulations or historical data to infer reward structures indirectly. This approach analyzes optimal behavior demonstrated by expert systems or agents to deduce the underlying reward functions. Inverse RL is particularly advantageous for tasks with available high-quality simulations or data, as it reduces manual engineering by learning from observed behavior.

However, this approach requires accurate models or datasets and can be computationally demanding.

### VIII.4. LLMs

LLMs, such as GPT-based models, provide an approach to reward design by enabling natural language interaction, automated tuning, and dynamic, context-aware recommendations. LLMs allow users to input goals in plain language – such as "maximize resolution while minimizing artifacts" – and translate them into mathematical reward functions. These models can also refine and adapt rewards based on evolving constraints by analyzing task performance and providing real-time adjustments. Additionally, LLMs facilitate collaborative reward development through shared tools and open-source platforms, making the process accessible even to non-experts. While LLMs significantly reduce manual effort and enhance adaptability, their outputs require careful validation to ensure alignment with domain-specific constraints.

### VIII.5. Roll-out functions.

The intermediate rewards approximations are necessary for decision algorithms such as MCDT and A*. These have to satisfy certain requirements (e.g. be optimistic). Practically for imaging workflows, they should allow to estimate the true reward function at the intermediate steps. A possible approach are the variants of the PCA analysis with reward function being identified as maximal or weighted average of reward functions of several PCA components. However, this requires separate developments.

## IX. HAE workflows

Reward-based workflows are inherently Human-Augmented Experimentation (HAE) workflows, as they enable dynamic adjustments to reward structures during experiments, allowing researchers to guide the process in real time. This approach empowers researchers to modify objectives based on evolving needs, such as shifting rewards to prioritize one objective over the other, depending on experimental demands. For example, in LoG* workflows for atomic imaging as shown in **Figure 11**, a researcher might adjust rewards mid-experiment to emphasize precision in predicting atomic positions (meaning reduce True positives) or maximizing the number of predictions (meaning increase False positives).[190] Additionally, HAE workflows allow researchers to interact with Pareto fronts in multi-objective optimization, selecting trade-offs that best align with the overarching experimental goals, such as balancing quality and quantity. This human-in-the-loop approach integrates human judgment and flexibility with automated workflows, enabling adaptable, efficient, and context-aware experimentation.

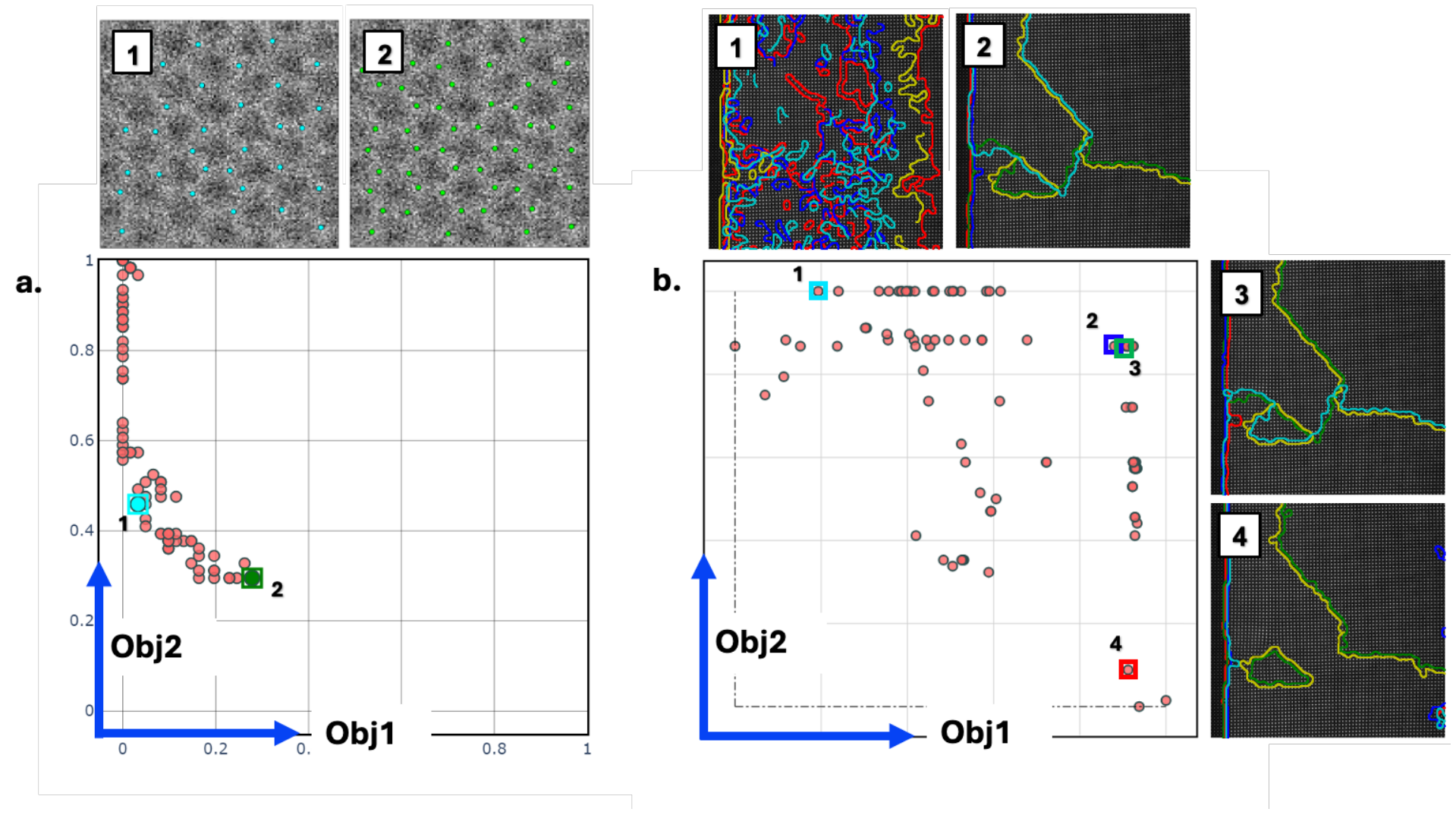

**Figure 11:** Human-Augmented Experimentation (HAE) in Reward-Based Workflows. (a) Pareto front solutions in V.1. LoG* workflow for atom finding, where *a[1]* and *a[2]* illustrate how reward adjustments shift priority between objective_1 and objective_2 for real-time optimization, and b) Pareto front solutions in V.3. Ferroelectric domains workflow, with *b[1], b[2]*, *b[3]* and *b[4]* similarly demonstrating objective prioritization through reward tuning. (Reproduced with pending permission from Barakati et al.)[190, 195]

The same concept also applies to automated optimization of microscopes. The optimization of microscopes often involves the balance between scan speed, reduction of damage to the samples, and the image quality. With rewards defined for each of these factors, HAE workflows allow the users to choose the desired trade-offs from the pareto front to achieve their experimental goals. These strategies can be applied to the multi-objective structure-property discovery experiments such as Deep Kernel Learning (DKL)[212], enabling the identification of scientifically valuable data trajectories while avoiding local minima.[213, 214] By translating EDX, EELS, and diffraction signals into robust reward functions, this strategy has the potential to enhance STEM and SPM data acquisition.

## Summary

The shift toward automated experiments and faster image analysis has made reproducibility and traceability more important than ever. This transformation requires building and optimizing workflows that address how early decisions can heavily influence later outcomes, much like natural language processing tasks where raw data evolves into richer, more meaningful representations. Human-driven analytics, whether consciously or not, often rely on reward functions that capture expert knowledge, derived either from physics-based principles or heuristics, to refine best practices. Once a reward function is defined, workflows can be optimized systematically, though challenges arise in processes like clustering, which are complex and non-differentiable.

The engineering of reward functions should be considered as an independent scientific challenge rather than a purely technical task. Effective reward construction requires integrating human heuristics, domain-specific physical insight, and the constraints of experimental systems. In real scientific multistep workflows, oversimplified rewards often fail to capture the complexity of the processes, leading to suboptimal convergence. As a result, reward design frequently requires probabilistic and multi-objective formulations. Designing reward-based workflows involves constructing data structures that accurately represent the sequence of operations, enabling them to be tackled using optimization strategies like genetic algorithms or classical decision-making frameworks. Algorithms like A* or Monte Carlo Decision Trees additionally require well-defined rollout functions to predict rewards effectively. While automation holds promise, expert input remains essential, whether through direct involvement or innovative tools like large language models or crowdsourcing to refine reward metrics.

The potential impact of reward driven workflow design is the transition from the supervised black box neural network and biased non-myopic human based analysis to the unsupervised explainable methods. By integrating advanced algorithms, reinforcement learning, and domain-specific expertise, we map the image analysis problem to decision-making process and allow image analytics to yield semantically-rich representations tailored to provide insight into specific questions. This shift also democratizes access to cutting-edge analytics, enabling researchers with less computational experience to focus on interpretation and discovery rather than manual optimization and enable human- and ML-based decision-making in autonomous experiments.

**Acknowledgement:**

This work (workflow development, reward-driven concept) was supported (K.B., Y.L., S.V.K.) by the U.S. Department of Energy, Office of Science, Office of Basic Energy Sciences as part of the Energy Frontier Research Centers program: CSSAS-The Center for the Science of Synthesis Across Scales under award number DE-SC0019288. The work was partially supported (U.P.) by AI Tennessee Initiative at University of Tennessee Knoxville (UTK). The authors particularly acknowledge Miles Cranmer (Cambridge), whose PySR system have inspired many of these developments, Jurafsky and Martin, the authors of "the Speech and Language Processing",[215] Peter Norvig and Russell, the authors "Artificial Intelligence: A Modern Approach"[216] and Alaa Kahmis, the author "Optimization Algorithms: AI techniques for design, planning, and control problems,[217] whose work inspired the concepts proposed here.

**DATA AVAILABILITY:**

The code supporting the findings of this study is publicly accessible on GitHub at [**GithuB**]. Additional methodological details and reproduction instructions for workflows in sections V.1–V.4 can be found in the referenced prior publications190, 194-196, which provide full technical descriptions, optimization procedures, and dataset information.